\documentclass[10pt, twocolumn]{article}

\usepackage[show]{chato-notes}

\usepackage{xr-hyper} %

\usepackage{pdfpages}
\usepackage{type1cm} %
\usepackage{xcolor}
\usepackage{graphicx}
\usepackage{amsmath,amssymb,amsthm}
\usepackage{mathtools} %
\usepackage{bbm}
\usepackage{booktabs}
\usepackage{epsfig}
\usepackage{url}
\usepackage{hyperref}
\usepackage{color}
\usepackage{fullpage}
\usepackage{tikz}
\usepackage{placeins}
\usetikzlibrary{bayesnet}
\usepackage{algorithm}
\usepackage{algpseudocode}
\usepackage{cleveref}
\usepackage{xspace}
\usepackage{balance} %
\usepackage{multirow} %
\usepackage[font=footnotesize, tableposition=top]{caption} %
\usepackage{caption}
\usepackage{subcaption}
\usepackage{siunitx} %
\usepackage{microtype} %
\usepackage{xfrac} %
\usepackage[square,numbers]{natbib} %
\usepackage[affil-it]{authblk}

\usepackage[small]{titlesec}

\makeatletter
\long\def\XR@test#1#2#3#4\XR@{%
  \let\XR@next\@gobbletwo
  \ifx#1\newlabel
    \let\XR@next\@firstoftwo%
  \else\ifx#1\@input
     \let\XR@next\@secondoftwo
  \fi\fi
   \XR@next{\newlabel{\XR@prefix#2}{#3}}{\edef\XR@list{\XR@list#2\relax}}%
  \ifeof\@inputcheck\expandafter\XR@aux
  \else\expandafter\XR@read\fi}
\makeatother

\hypersetup{
    bookmarks=true,         %
    unicode=false,          %
    pdftoolbar=true,        %
    pdfmenubar=true,        %
    pdffitwindow=false,     %
    pdfstartview={FitH},    %
    pdftitle={My title},    %
    pdfauthor={Author},     %
    pdfsubject={Subject},   %
    pdfcreator={Creator},   %
    pdfproducer={Producer}, %
    pdfnewwindow=true,      %
    colorlinks=true,       %
    linkcolor={red!50!black},
    filecolor={green!50!black},
    citecolor={green!50!black},
    urlcolor={blue!80!black},
}

\newcommand{\indep}{\perp \!\!\! \perp}
\newcommand{\prob}{\mathbb{P}}

\newtheorem{principle}{Principle}
\crefname{principle}{Principle}{Principles}

\algrenewcommand\algorithmicrequire{\textbf{Input:}}
\algrenewcommand\algorithmicensure{\textbf{Output:}}

\newenvironment{squishlist}
{\begin{list}{$\bullet$}
 {\setlength{\itemsep}{0pt}
     \setlength{\parsep}{3pt}
     \setlength{\topsep}{3pt}
     \setlength{\partopsep}{0pt}
     \setlength{\leftmargin}{1.5em}
     \setlength{\labelwidth}{1em}
     \setlength{\labelsep}{0.5em} } }
{\end{list}}

\graphicspath{{./figures/}{../plots/}}

\newcommand{\dropcap}[1]{#1} %

\title{On learning agent-based models from data}
\author[a,1,2]{Corrado Monti}
\author[b,a,1,2]{Marco Pangallo}
\author[a,2]{Gianmarco De Francisci Morales}
\author[a,2]{Francesco~Bonchi}

\affil[a]{CENTAI Institute, Turin, Italy}
\affil[b]{Institute of Economics and EMbeDS department, Sant'Anna School of Advanced Studies, Pisa, Italy}

\date{\today}

\begin{document}

\twocolumn[
  \begin{@twocolumnfalse}
    \maketitle
    \begin{abstract}
      Agent-Based Models (ABMs) are used in several fields to study the evolution of complex systems from micro-level assumptions.
However, ABMs typically can not estimate agent-specific (or ``micro'') variables: this is a major limitation that prevents ABMs from harnessing micro-level data availability and which reduces their predictive power.
In this paper, we propose a protocol to learn the latent micro-variables of an ABM from data.
The first step of our protocol is to reduce an ABM to a probabilistic model, characterized by a computationally tractable likelihood.
Then, we proceed by maximizing the likelihood of the latent variables via a gradient-based expectation maximization algorithm.
We showcase our protocol by applying it to an ABM of the housing market, in which agents with different incomes bid higher prices to live in high-income neighborhoods.
We demonstrate that the obtained model produces accurate estimates of the latent variables, while preserving the general behavior of the ABM. 
We also show that our estimates substantially improve the out-of-sample forecasting capabilities of the ABM compared to simpler heuristics.
Our protocol can be seen as an alternative to black-box data assimilation methods,
that forces the modeler to lay bare the assumptions of the model, to think about the inferential process, and to spot potential identification problems.

    \end{abstract}
    \vspace{1em}
  \end{@twocolumnfalse}
]

\footnotetext[1]{These authors contributed equally to this work.}
\footnotetext[2]{To whom correspondence should be addressed. E-mail: corrado.monti@centai.eu; marcopangallo@gmail.com; gdfm@acm.org; bonchi@centai.eu}

\dropcap{A}gent-Based Models (ABMs) are computational models in which autonomous ``agents'' interact with one another and with their environment, thereby producing aggregate emergent phenomena \citep{wilensky2015introduction}.
ABMs are an extremely successful tool for theory development, that is, to explore the macro-level implications of micro-level assumptions~\citep{railsback2019agent}.
As Axelrod \cite{axelrod1997dissemination} said \emph{``whereas the purpose of induction is to find patterns in data [...], the purpose of agent-based modeling is to aid intuition''}.
In line with this focus on theory development, the ability of ABMs to match empirical data and make quantitative forecasts---that is, to \emph{learn from data}---has been, so far,  limited~\citep{lux2018estimation,gatti2020rising,windrum2007empirical,deffuant2008agent}.

\begin{figure*}[!h]
\begin{center}
\includegraphics[width=1\textwidth]{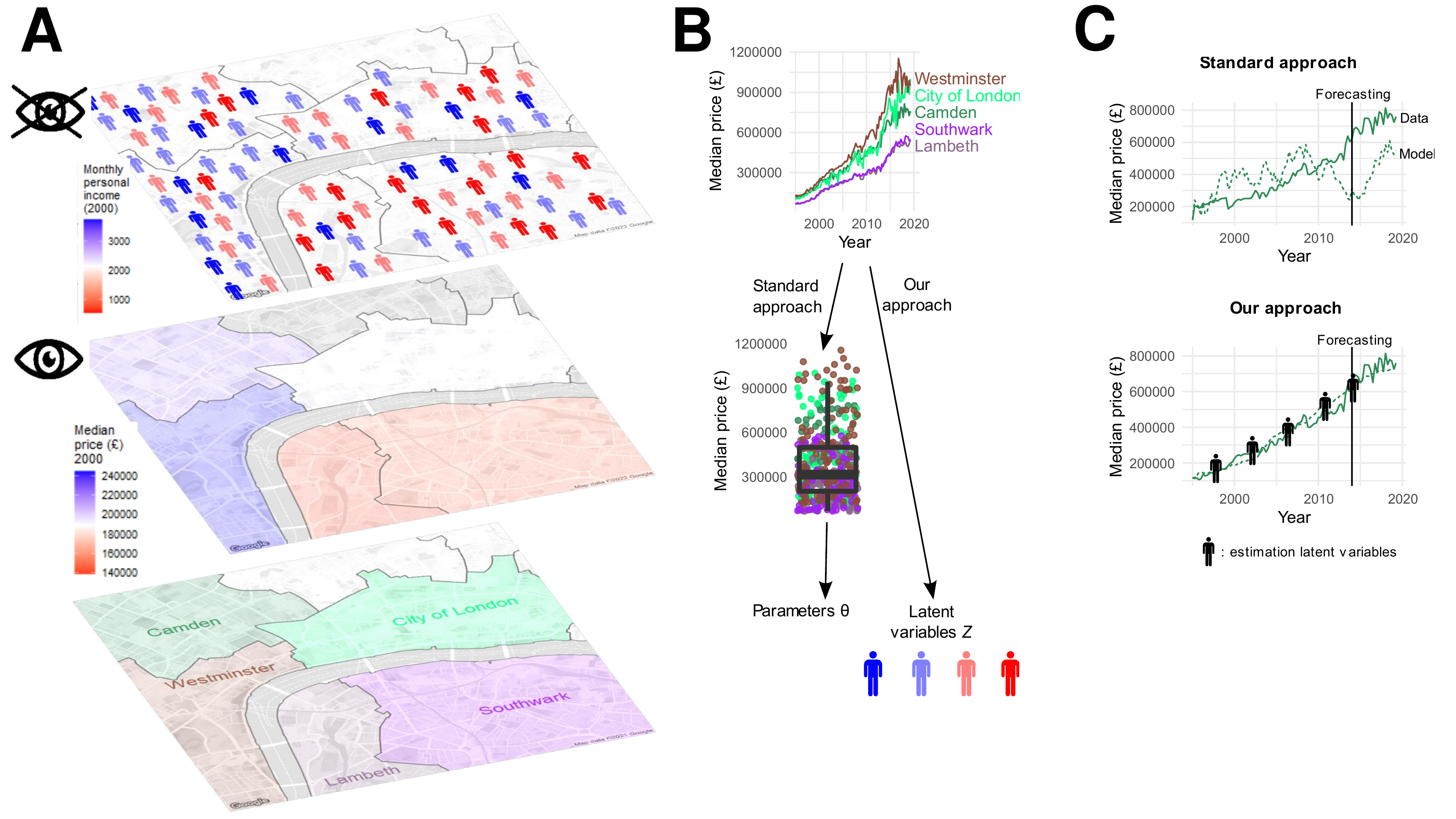}
\caption{\textbf{Our approach compared to a standard approach towards calibrating an ABM of the housing market.} (A) Focusing on the boroughs in the center of London (bottom layer), we consider the yearly average of transaction prices (middle layer) as an example of an observed variable, and the distribution of agent incomes (top layer) as an example of a latent variable. (B) For each borough, we observe a time series of transaction prices. In the standard approach to calibration, modelers could typically calibrate some parameters  $\Theta$ (such as the probability for inhabitants to put their home on sale) by computing the moments of transaction prices across boroughs and years (as represented through a box plot), and minimizing the distance with the same moments in model-generated time series. In our approach, instead, we are able to calibrate the evolution of latent variables $Z$ --in this example, borough-level agent incomes-- by exploiting all information contained in time series, rather than reducing this information to specific summary statistics. (C) In the model, prices depend on agent incomes. Thus, since in the standard approach agent incomes are not calibrated, model-generated time series are bound to diverge, even if prices are initialized as in the data. With our approach, as we repeatedly estimate incomes, we can make model-generated time series track empirical ones. This makes it feasible to forecast future prices.}
\label{fig:schematic_intro}
\end{center}
\vspace{-\baselineskip}
\end{figure*}

At a very high level, all ABMs can be described by the formula 
\begin{align}
Z_t \sim \prob_t ( Z_t \mid \Theta, Z_{\tau < t} ) ,
\end{align}
where $Z_t$ are the \emph{variables} of interest in the system, $\Theta$ is a set of \emph{parameters}, $\prob_t$ is a probability function implicitly defined by model specifications, and $t$ is the discrete time index.
Typically, $\Theta$ has relatively few components and a fixed dimensionality, is interpretable by a domain scientist, and is the only tuning knob of the model.
$Z$ is at the same time the state and the output of the model.
Each component of $Z$ typically refers to an individual agent, which results in high dimensionality.
Some of $Z$ is \emph{observable}, while the rest is \emph{latent}.

Most of the efforts in learning agent-based models from data have focused on parameter calibration. 
This task refers to the process of finding a parametrization $\Theta$ that can reproduce some macroscopic characteristic of the data, and it typically boils down to comparing a few summary statistics of aggregate empirical and model-generated variables (e.g., time series)~\citep{windrum2007empirical,lee2015complexities}. 
Summary statistics are valuable to focus on the most important characteristics of the data that the modeler wants to explain, but they often have to be chosen arbitrarily, and may hide very different underlying patterns (as in the well-known Anscombe quartet). 

Estimating agent-specific (or ``micro'') \emph{variables} $Z$ is instead not usually considered.
We think this is the main obstacle to bringing ABMs closer to data and potentially using them as a forecasting tool.
Indeed, if modelers do not correctly initialize latent micro variables and update their values as the simulation progresses, and if the dynamics of observed variables crucially depend on the initialization of latent variables, then model-generated time series are bound to diverge from empirical ones.
Additionally, this mismatch implies that model-generated and empirical time series cannot be directly compared to produce a ``goodness-of-fit'' measure, so one must resort to summary statistics or ``stylized facts'' to calibrate parameters.
The ABM community has recently started to explore data assimilation methods to estimate the latent variables of ABMs \cite{ward2016dynamic,lux2018estimation,clay2021real,cocucci2022inference}; we explain the relation of this literature to our work in the Discussion.

This paper proposes a general methodology for estimating latent variables in ABMs.
Our approach proceeds in three steps:
\begin{enumerate}
\item Given an ABM of reference, translate it into a probabilistic model, by simplifying it until a computationally tractable likelihood of the data given the latent variables can be specified.
\item Estimate latent variables at each time step while keeping all past input values fixed (as in online learning).
Solidly rooted in probability theory, our approach maximizes the likelihood at each step via expectation-maximization~\citep{dempster1977maximum} and gradient descent.
\item Repeat this process over multiple epochs, so that temporal dependencies can be appropriately taken into account.
\end{enumerate}

We showcase our approach by applying it to a housing market ABM specifically designed to study income segregation \citep{pangallo2019residential} (\Cref{fig:schematic_intro}).
We use the ABM to generate synthetic data traces that we can use as ground truth (we would not have access to a ground truth for latent variables if we used real-world data traces).
We distinguish between observable and latent variables based on how easy it usually is to access the relevant real-world data.
The main latent variable in this model is the distribution of agents' incomes in each neighborhood, which is often not available.
We instead assume that we observe neighborhood-level mean prices and the number of transactions over time (these data are usually readily available, see e.g. \cite{loberto2022whatdo}).
We believe this distinction to reflect a common real-world setting in economic models: one might have access to which actions are being performed, but not to the latent state of agents (e.g., where they live, and what is their income).
We write the likelihood of prices and of the number of transactions as a function of the household income distribution.
Next, we maximize this likelihood, thus estimating the time evolution of the spatial income distribution.

In synthetic experiments, we show that our procedure enables learning latent variables accurately: the Pearson correlation between the ground truth and learned traces ranges between 0.5 and 0.9, depending on the latent variable considered.
At the same time, we show that an accurate estimation of latent variables empowers out-of-sample forecasting.
Compared to other benchmarks that use rules of thumb to initialize the model at the beginning of the forecasting window, our procedure obtains lower Root Mean Squared Error (RMSE) with respect to the ground truth while being more principled.
It also highlights potential identification problems, i.e., situations wherein multiple configurations of micro-variables correspond to the global maximum of the likelihood, so that the ground truth configuration cannot be identified.
\vspace{-0.5\baselineskip}

\section*{Model}
\label{sec:model}

We start from the housing market ABM presented by Pangallo et al.~\cite{pangallo2019residential}---henceforth, the ``original ABM''.
Our goal is to modify the original ABM until it is possible to write a computationally tractable likelihood of observed variables given latent variables and parameters.
If we are able to do so, we say that the modified model is a \emph{learnable} ABM.
While writing a tractable likelihood function, we need to preserve the general behavior of the model, as well as its essential causality mechanisms.

In this section, we first give an overview of the original ABM; then, we describe the learnable ABM resulting
from our `translation' process. Along the way, we highlight the specific transformations needed to make the original ABM learnable.
A more detailed explanation of the equations describing our learnable model is given in Materials and Methods and summarized in \Cref{tab:model-equations}.
\Cref{fig:graphical-model} reports the causal links between variables as a graphical model.
Supplementary \Cref{sec:old-model-comparison} provides more details on the models and the trasnlation process.

\paragraph{Original ABM \cite{pangallo2019residential}.}
The ABM describes the housing market of a city composed of $L$ locations or neighborhoods, each with a number of indistinguishable homes, inhabited by agents.
Each agent belongs to an income class $k$, out of $K$ income classes, each characterized by an income $Y_k$.
At each time step, individual agents---represented as discrete units---choose a neighborhood to purchase a home if they act as buyers, or put their home on sale if they act as sellers.
One fundamental insight encapsulated in this model is the formalization of the \emph{attractiveness} of each neighborhood, which regulates how likely an agent is to bid for that location.
The model assumes that the higher the income of residents, the more attractive a neighborhood is.
In this original model, matching between individual buyers searching in a neighborhood and sellers in the same neighborhood is modeled as a \emph{continuous double auction}.
This process selects buyers and sellers sequentially at random, puts buyers in a queue ordered from highest to lowest bid price (and sellers from lowest to highest ask price), and, whenever a seller asks a price below the maximum bid price in the queue, matches the buyer with highest bid price to the seller with the lowest ask price.
The social composition of the city evolves as a byproduct of these transactions, as high-income buyers may replace low-income sellers and lead to the gentrification of some neighborhoods.
We report the pseudocode of this original model in \Cref{alg:full-abm-run}.

\paragraph{Learnable ABM.}
In order to translate such ABM into a learnable model, we first rewrite it in terms of `counts', i.e., instead of having variables for each individual agent, with a small loss of generality we consider the aggregated information about the number of identical agents of each income class in each location.
This way, we obtain a model that revolves around the state variable $M_t$:
at each time step, $M_t$ is a matrix of $L \times K$ entries, where $M_{t, x, k}$ represents the number of agents of income class $k$ in location $x$ at time $t$.
Similarly, the number of agents of class $k$ buying a house in location $x$ is represented by $D^B_{t,x,k}$, giving a total of $D_{t, x}=\sum_k D^B_{t,x,k}$ transactions.
$D_{t, x}$ is in turn determined as the short side of the market, i.e., the minimum between the number of sellers and the number of buyers in each case. 
While these two numbers in the original model were stochastic, in our learnable model we use a mean field approximation, and replace the stochastic realizations with their expected value.
The final key variable is $P_{t, x}$, which represents the average price of transactions that occur in location $x$ at time $t$.

\paragraph{Matching protocol.}
The matching protocol between buyers and sellers clearly exemplifies the type of transformations needed for our purpose.
The continuous double auction of the original ABM is indeed hard to translate into a computationally tractable likelihood.
First, we assume that we do not have detailed information on buyers and sellers for individual transactions, so estimating, e.g., the stochastic sequence in which buyers and sellers enter the queue is not feasible.
Second, picking the buyer with highest proposed price is equivalent to an $\operatorname{argmax}$ operation.
Such operation is not differentiable, thus causing the whole likelihood to be not differentiable.
Indeed, estimating its outcome would require enumerating all possible cases.
To solve both issues, while preserving the properties of the model, we replace the continuous double auction by a multinomial distribution that gives higher probability of matching to buyers proposing higher prices (Equation \ref{eq:M8} in Table \ref{tab:model-equations}).
This rule is differentiable and can be estimated from observed prices: higher prices indicate that richer agents have settled in the neighborhood.

\section*{Algorithms}
\label{sec:algorithms}

Once we have translated the original ABM into its learnable counterpart, we design an algorithm that infers latent variables by maximizing the likelihood of these variables with respect to observed data and model's assumptions.

To start with, we need to determine which variables are observed and which are latent.
To do so, we think of aggregate information about transactions as the only observable at our disposal.
In particular, we assume to know, for each neighborhood and over time, the number of transactions $D_t$ and the average price $P_t$.
Our key latent variable is instead $M_t$, the distribution of agents of each income class across neighborhoods.
We believe this distinction to reflect a common real-world setting in economic models: one might have access to which actions are being performed, but not to the latent state of agents (e.g., where they live, and what is their income).
As a matter of fact, in many countries it is relatively easy to obtain spatially granular data on transactions, but it is much harder to obtain such data on incomes \cite{loberto2022whatdo}.

Note that $M_t$ can be computed deterministically given $M_{t-1}$ and $D^B_{t-1}$.
Therefore, our problem reduces to finding an estimate for the latent stochastic variable $D^B_t$, over all time steps $t=1,\ldots,T$, and for the starting condition $M_0$, given $P_t$ and $D_t$:
all the other variables are in fact deterministic, and their value is fixed given the formers.
This scenario corresponds to the graphical model shown in \Cref{fig:graphical-model}.

However, the number of possible states of $D^B$ grows exponentially with the total number of time steps $T$: evaluating all possible paths of agents over all time steps would be unfeasible even for small values of $T$.
Therefore, we approach our problem as an online task~\cite{monti2020learning}, a common technique in machine learning in cases where processing the entire data set at once is unfeasible.
We process the data per time step:
at each $t$, the algorithm is presented with the newly observed values $D_t$ and $P_t$, and it updates its estimate of the latent variables $M_0$ and $D^B_t$, while considering all the values previously estimated as fixed.
After the given time step $t$ has been processed, the algorithm is applied on $t+1$, and so on until the last time step $T$.
This process---examining each time step from $t=0$ to $T$---is iterated for a number of epochs: after the last time step $T$ has been processed, the algorithm re-starts from the beginning, so that the first time steps are re-evaluated in light of successive ones.

To solve this optimization problem, we propose an expectation-maximization algorithm.
Such an algorithm is able to obtain a maximum-likelihood estimate of the latent variables by optimizing the complete-data likelihood of the model.
We outline its derivation in Materials \& Methods \Cref{sec:likelihood}.
It operates by repeating at each given time step $t$ the following process.
First, it evaluates the likelihood of each possible behavior of the agents---i.e, the possible outcomes of $D^B_t$; then, it uses back-propagation and online gradient descent to find the best $M_0$ under this likelihood.
These two steps are alternated until convergence.
This way, at each time step it recovers the most likely value for $D^B_t$, and it updates its estimate for $M_0$.
All the other variables of the ABM are obtained deterministically from these ones.

In order for the algorithm described thus far to be scalable, we need to solve one last computational challenge:
even in a single time step $t$, the space of possible outcomes of each $D^B_t$ is huge since in principle one should consider the decisions of all individual buyers as independent. 
We solve this problem by considering that, while in agent-based modeling it is common to model the behavior of individual agents, for our purposes is sufficient to evaluate the chances of \emph{groups} of identical agents moving to one location or another: the behavior of a single agent is irrelevant with respect to the data we observe.
Therefore, instead of considering all the possible outcomes of each $D^B_t$, we consider only those set apart by at least $s$ agent, where $s$ is a learning hyper-parameter.

\section*{Experiments}
\label{sec:experiments}

In order to evaluate the efficacy of our approach, we perform two sets of experiments.
First, we assess its fidelity, i.e., how well our method recovers latent variables.
To do so, we generate a synthetic dataset from the original ABM as ground truth, and feed the observable part of such data to our likelihood-maximization algorithm.
Second, we show that learning latent variables allows us to produce more accurate out-of-sample forecasts compared to existing heuristics.

\subsection*{Recovering latent variables}

We consider the time series of the price $P_{t,x}$ and of the number of transactions $D_{t,x}$ at each location $x$ to be observable.
We also assume that the macro-parameters that generated the data set (e.g., the total number of agents per location $N$, or the global income distribution) are known.
Two other time series, namely that of inhabitants $M_{t,x,k}$ and buyers $D^B_{t,x,k}$, for all locations $x$ and incomes $k$, are considered latent:
they are hidden from the algorithm and used as a validation for what the algorithm learns.

\begin{figure}[t!]
  \centering
  \begin{subfigure}[b]{1.0\linewidth}
      \centering
      \includegraphics[width=1.0\linewidth]{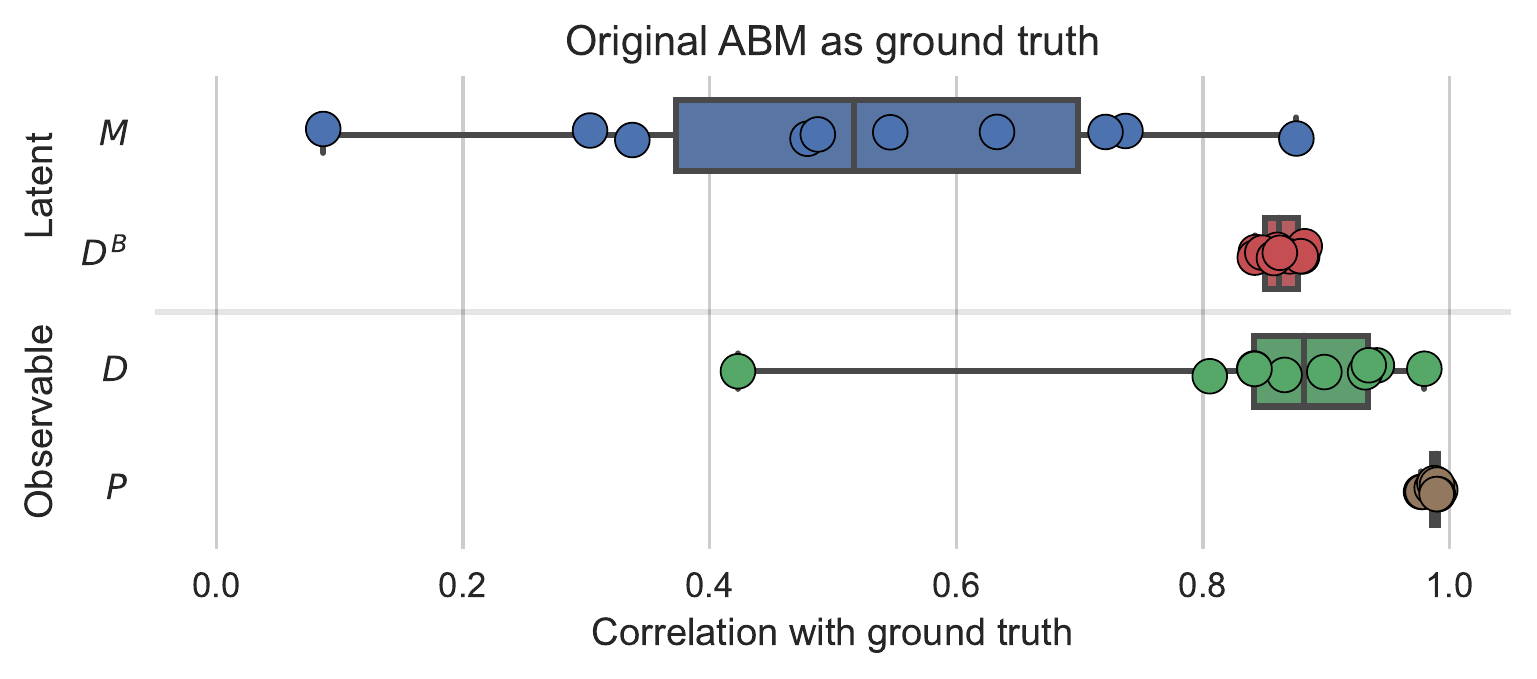}
  \end{subfigure}
  \vspace{2mm} \\
  \begin{subfigure}[b]{1.0\linewidth}
      \centering
      \includegraphics[width=1.0\linewidth]{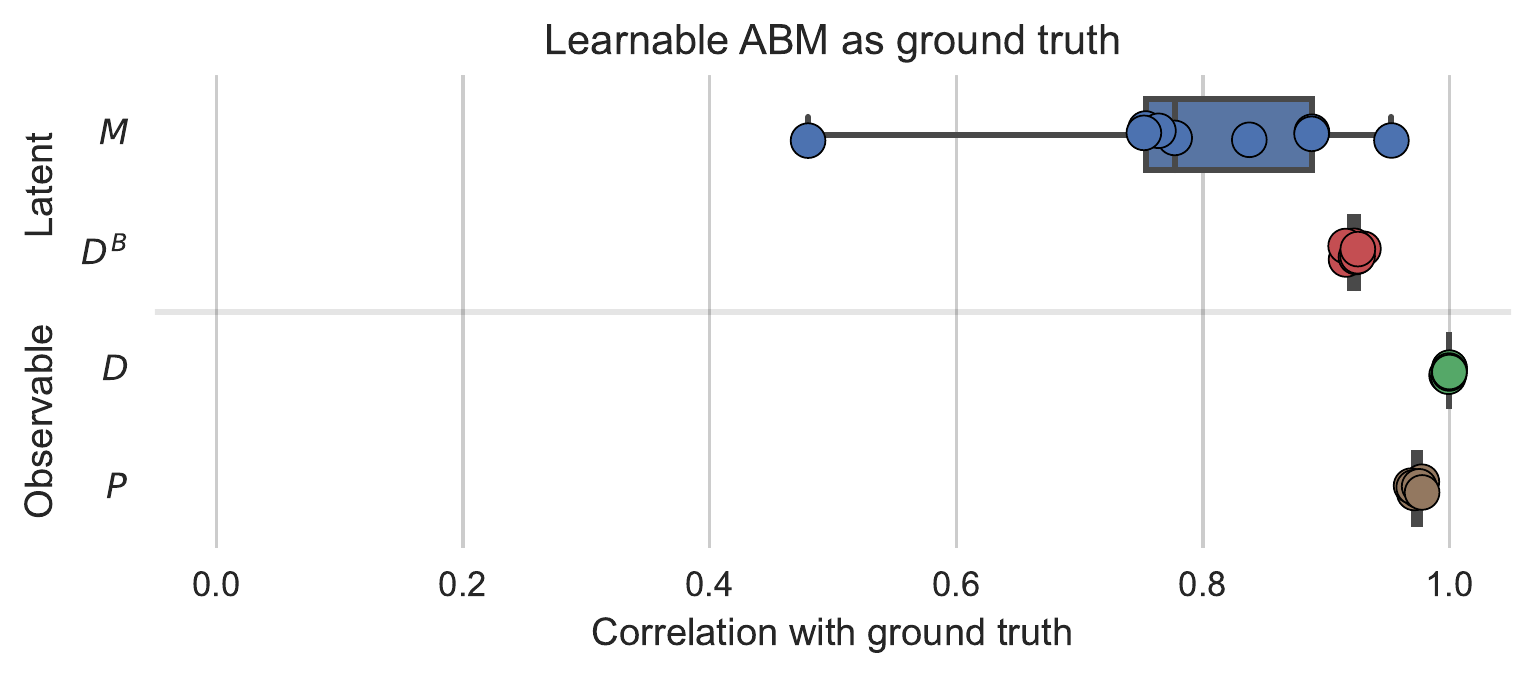}
  \end{subfigure}
  \caption{\label{fig:estimates} %
  Quality of estimation in synthetic experiments with traces generated by the original (top) and learnable (bottom) ABM. %
  For each variable, we report the Pearson correlation between the original values and the estimates.
  We represent each trace as a dot, with whisker plot as a summary for each variable.
  Whiskers extend from the minimum to the maximum value, while boxes range from the 25th to the 75th percentile.
  }
\end{figure}

We use the original ABM to generate 20 data traces with $L=5$ locations, $K=3$ income classes, and $T=20$ time steps that we use as ground truth.
Each data trace differs from the others in the random initialization of $M_0$.
We use the first 10 traces as training set to tune the hyperparameters of the algorithm (see Supplementary \Cref{fig:hyperparam-pearson}).
Then, on the remaining 10 traces that we hold out as test set, we evaluate the performance of the algorithm by computing the Pearson correlation coefficients between learned time series and ground truth ones.
Note that our learning algorithm uses the data from the learnable ABM to specify the likelihood, so there is some misspecification compared to the original ABM used to generate the ground truth.
For completeness, we also repeat the same evaluation by using the learnable ABM to generate ground-truth data traces, thereby removing misspecification.
We view this latter test as a sanity check for the algorithm.

\Cref{fig:estimates} shows the results for the test sets.
As expected, our ability to reconstruct latent variables is higher for traces generated with the learnable ABM, as there is no misspecification.
Perhaps more interestingly, our algorithm reconstructs the time series of buyers $D^B_{t,x,k}$ with higher fidelity than the time series of inhabitants $M_{t,x,k}$ (mean correlation $\rho=0.86$ vs. $\rho=0.52$ with traces from the original ABM).
Even though $M_{t,x,k}$ proves to be harder to reconstruct, we still obtain an informative estimate, that further improves when misspecification is removed ($\rho=0.79$).
We conjecture that this difference in results may due to the fact that $D^B_{t,x,k}$ is a ``flow'' variable that does not depend explicitly on previous time steps, while $M_{t,x,k}$ is a ``stock'' variable that depends on the whole history, so errors in estimating $M_{t=0}$ accumulate at following time steps.
These results also hint at an identification problem in the original ABM, which we elaborate on further at the end of this section.
Regarding the observable variables, our algorithm fits the prices $P_{t,x}$ almost perfectly ($\rho=0.99$), and the number of transactions $D_{t,x}$ very well ($\rho=0.85$).
Without misspecification, the fit for the number of transactions is perfect ($\rho=1.0$).
While such a good fit for observable variables is expected, since our inference method works by minimizing the distance from observable variables, this result indicates that there is no major misspecification introduced by using the learnable ABM to infer latent variables from traces of the original ABM.
In other words, our translation does not alter the nature of the original model.

\begin{figure}[t!]
  \centering
  \includegraphics[width=1.0\linewidth]{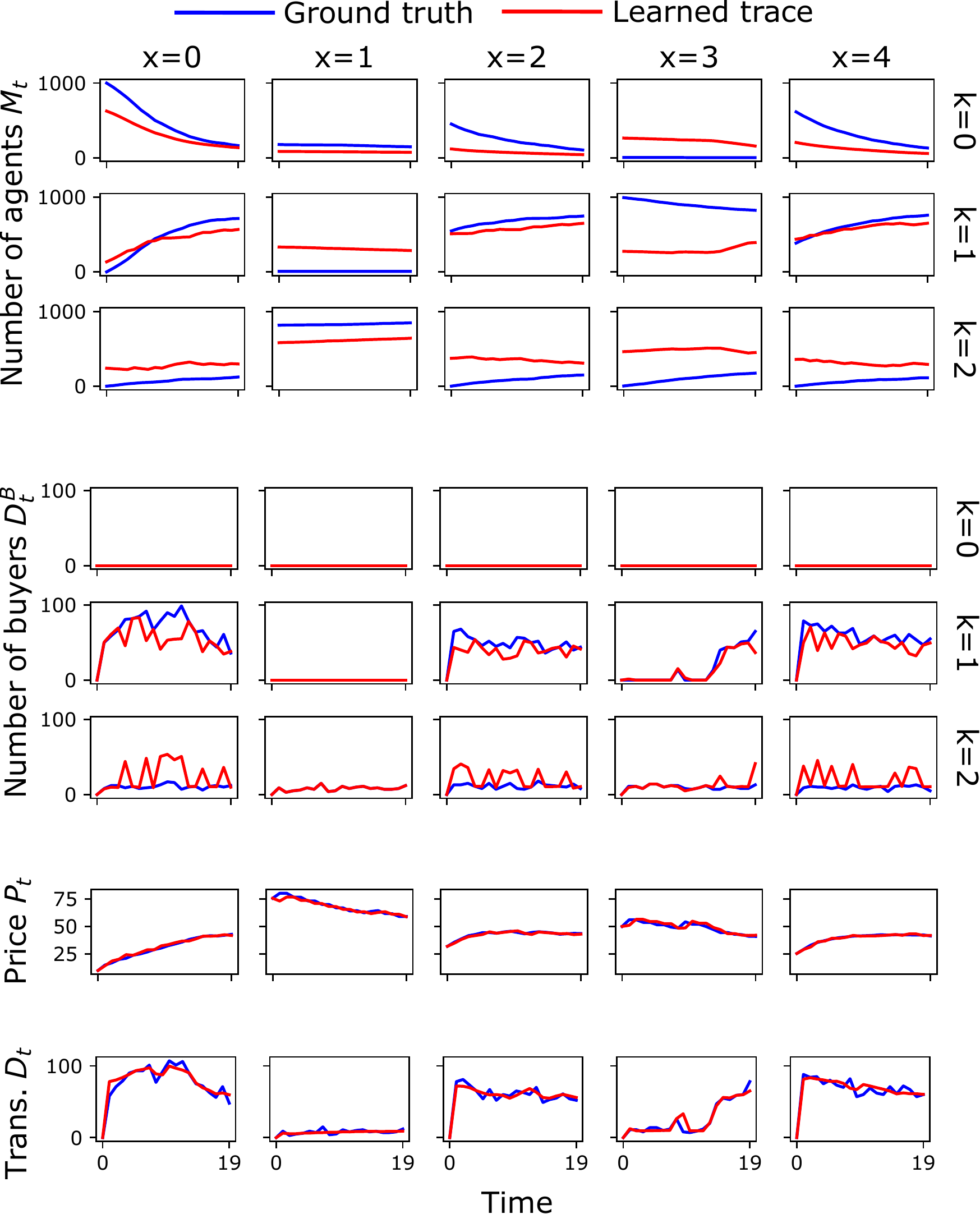}
  \caption{\label{fig:population-class} %
  Estimates for $M_{t,x,k}$, $D^B_{t,x,k}$, $P_{t,x}$, $D_{t,x}$ compared to the traces generated with the original ABM, in a single experiment, chosen as the median experiment in terms of estimation quality.}
\end{figure}

\Cref{fig:population-class} zooms in on a representative trace generated by the original ABM.
For fairness, we choose this experiment as the median one in terms of performance (i.e., correlation between the ground truth and estimated values of $M$).
These time series confirm the intuition from \Cref{fig:estimates}: our approach is able to reconstruct $P_t$ and $D_t$ extremely well and it is also quite precise at reconstructing $D^B_t$.
Our estimate of $M_t$ is also very accurate in most cases, but imprecise estimates of the initial conditions $M_0$ lead in a few cases (for instance, in location $x=3$) to an inaccurate reconstruction.
In a few cases, in fact, the algorithm finds a local minimum that does not correspond to the ground truth.

One of the possible reasons behind this behavior is the presence of an \emph{identification problem}.
We show in fact that, in some cases, the likelihood of the observed data is the same for different possible values of the latent variable $M_0$.
While these possible values include the ground truth (or, in case of misspecification, values extremely close to it) the model does not have enough information to distinguish it from the other possible optimal values of $M_0$.
This phenomenon is intrinsic to the ABM under study, once we identify $P$ and $D$ as observable and $M_0$ as latent.
We provide a concrete example in Supplementary~\Cref{sec:loss}.
\Cref{fig:heatmap-2-locs} shows a representation of the likelihood able to efficiently visualize such issues.
Our approach allows in fact to formally define and thus diagnose such issues.
Of course, one could also do this by sampling from the parameter space and computing summary statistics, as with Approximate Bayesian Computation (ABC) calibration methods (see, e.g., Ref.~\cite{dyer2021approximate}).
Our approach, which features a closed-form of the likelihood, has three advantages over these methods: ($i$) higher accuracy, as we do not have sampling error; ($ii$) higher efficiency, as we do not need to repeatedly execute the model; and ($iii$) the possibility to look for local minima using gradient-based methods.

\subsection*{Out-of-sample forecasting}

Except for a few recent attempts \cite{venkatramanan2018using,gatti2020rising,poledna2020economic}, so far the use of ABMs for forecasting has been limited. 
A key problem is that ABM state variables are mostly latent, as it is often hard to observe information that describes individual agents.
To the extent that the aggregate dynamics depend on the agent states, a wrong initialization of the latent state variables is likely to lead to a very inaccurate forecast.
In this section, we explicitly test whether this is true for our model by using synthetic experiments.
To shift our focus away from misspecification errors, we use the learnable ABM to generate the ground truth.
We extend each of the 10 test traces for 5 additional time steps, so that the total length of each simulation becomes $T'=25$.
To perform the forecast, we initialize the learnable ABM at $t=20$ with a given estimate of the latent state variables $M_{T=20}$, and let it produce the time series $P_t$ and $D_t$ for the out-of-sample time steps $t \in [21,25]$.
We compare five approaches for the estimate of the latent variable $M_T$:
\begin{enumerate}
\item \emph{Ground truth}: we use the true value of $M_T$ generated by the ABM, which we assume to be unobservable.
Because of inherent stochasticity, the forecast error is not zero.
However, this method represents a lower bound on the forecast error.
\item \emph{Random}: we draw $M_T$ from a Dirichlet distribution whose parameters are consistent with the share of buyers $\Gamma_k$. A random initialization of latent variables is very common in ABMs, for instance in epidemiological ABMs it is common to choose infected seeds at random \cite{aleta2020modelling}.
\item \emph{Proportional}: we draw $M_T$ in a way that locations with price higher than the mean price over the city have a higher share of high-income inhabitants.
The strength of this relation is governed by a hyperparameter that we calibrate in-sample on the same 10 traces that we use to select the hyperparameters of the learning algorithm.
\item \emph{Time series}: we run 1000 simulations starting from different values of $M_0$ and select the $M_T$ corresponding to the simulation with the lowest RMSE with respect to the observable time series $P_t$ and $D_t$ in sample, i.e., for $t \in [1,20]$. This is, for instance, the method used by Geanakoplos~et~al.~\cite{geanakoplos2012getting}.
\item \emph{Learnable}: we infer $M_T$ with our algorithm, by using the estimates obtained as specified in the previous section.
\end{enumerate}

\begin{figure}[t!]
  \centering
  \includegraphics[width=1.0\linewidth]{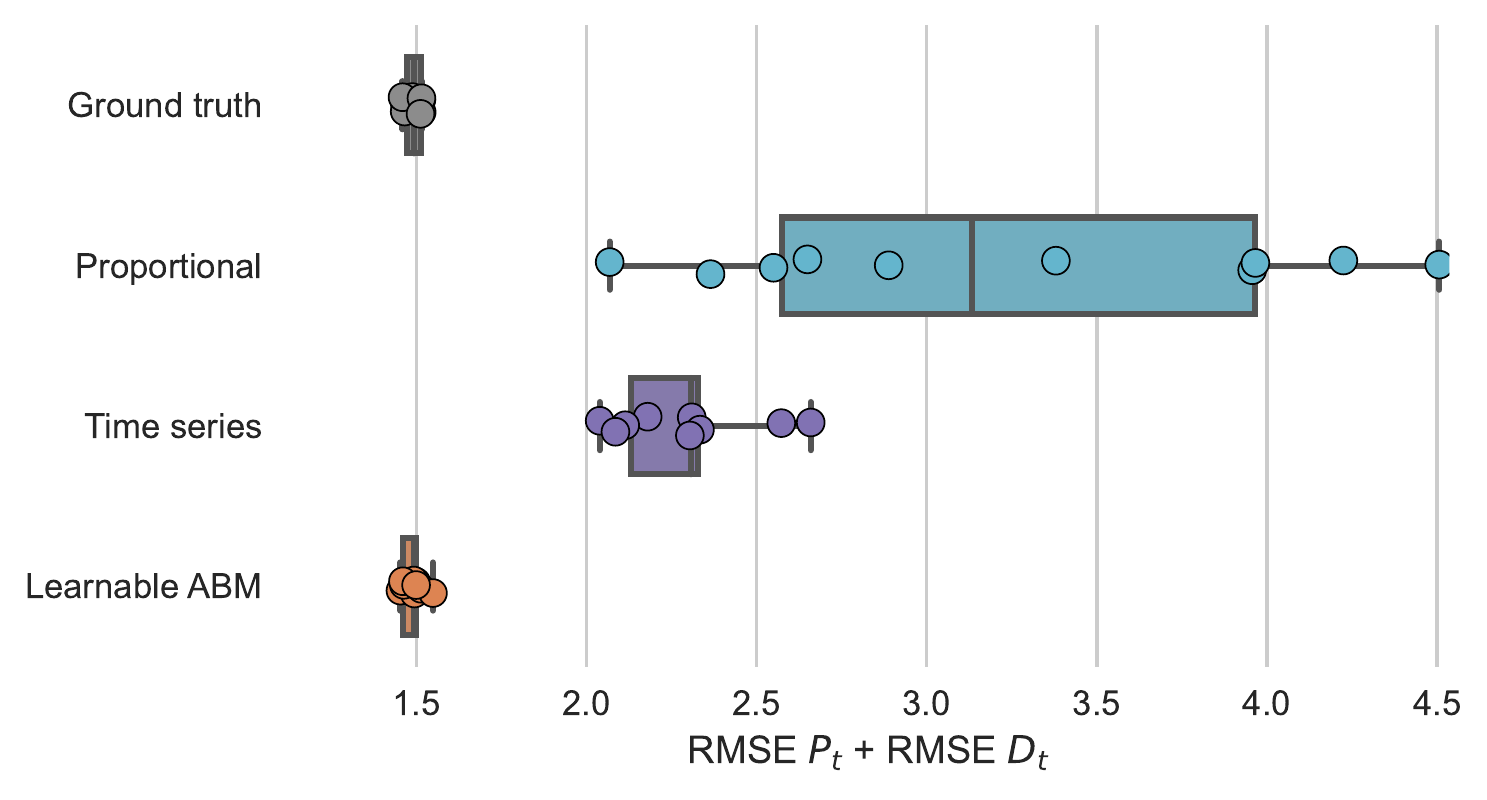}
  \caption{\label{fig:forecasting} Forecasting error for our method compared to alternative benchmarks. Forecasting error is measured as the sum of the RMSE on the $P_t$ and $D_t$ time series. We consider the same 10 traces as in the experiments above, and show results for each trace as a dot and a whisker plot as a summary. Whiskers extend from the minimum to the maximum value, while boxes range from the 25th to the 75th percentile.}
\end{figure}

To evaluate the quality of the forecasts obtained by these approaches, we compute the Root Mean Squared Error (RMSE) for the observable time series $P_t$ and $D_t$, summing the errors from time step $t=21$ up to $t=T'=25$.
\Cref{fig:forecasting} shows the results.
We do not report the values for the \emph{Random} approach as it has RMSE=12, well above that of the other approaches.
Our \emph{Learnable} approach substantially improves over the \emph{Proportional} and \emph{Time series} approaches, and essentially on par with the ground truth benchmark.
This is a strong improvement, but we believe that the value of our approach goes beyond this aspect.
In fact, alternative approaches are heuristics, that do not yield much insights about inference.
By contrast, our approach is more principled: it frames the problem of estimating unobservable variables of an ABM into probabilistic inference.
This methodology opens new research directions to further improve our results.
For instance, designing learnable ABMs from the start, for which there would be no misspecification error.
Even more importantly, it makes it possible to formally reason about the likelihood of an ABM---for instance, to spot potential identification problems.

\section*{Discussion}
\label{sec:discussion}

Now that we have shown the advantages of our method in terms of predictive capabilities, we discuss its general applicability.
From the specific translation of the housing market ABM considered in this paper, we can identify some general design principles, that we believe will be useful in making other ABMs learnable.

First of all, it was necessary to tune the level of stochasticity of the model by considering which variables are observed and which are latent.
In most graphical models, latent variables are stochastic random variables that are related to the observable ones---indeed, if they were deterministic, they could be computed exactly~\cite{welling2008deterministic}.
All stochastic variables that are not observed must be estimated, thus increasing both the computational complexity of the process and the uncertainty of the model.
However, in our translation, we have room to decide which variables are deterministic and which are stochastic.
To make the model truly learnable, we need to balance observable and latent variables so that for every latent variable we have some observable that intuitively makes it possible to estimate it. 
We can encapsulate this first design principle as follows.

\begin{principle}
  \label{principle:deterministic}
\textbf{Stochasticity parsimony.}
In a learnable ABM, the amount of stochasticity should be commensurate to data availability.
\end{principle}

\smallskip

Second, we needed to carefully consider which variables and functional forms should be discrete.
ABMs often consist of discrete units, and it is common for agents to choose between different discrete possibilities.
However, discreteness makes likelihood optimization problematic.
Indeed, whenever we deal with discrete variables, the likelihood must consider all possible combinations of values for discrete stochastic variables, which greatly increases the computational burden of the approach.
Moreover, in some cases the likelihood may be flat over some region of the latent space, thus hindering the progress of optimization algorithms such as gradient descent.
Given these considerations, it is important to limit the use of discrete variables to the ones that are critical to the behavior of the model.

\begin{principle}
  \label{principle:differentiable}
  \textbf{Differentiability preference.}
A learnable ABM should prefer continuously differentiable functions over discrete choices when they do not alter the behavior of the model.
\end{principle}

Following these principles, it should be possible to transform any ABM into a learnable one, given enough data.
While the translation in this paper was still hand-curated, it is a first step towards its proper formalization, and thus automatization.

\smallskip

Nevertheless, different alternative methods have been suggested in the literature to obtain similar results.
Making the state of the system compatible with real-world observations has traditionally been the goal of data assimilation techniques, such as the various versions of the Kalman filter or the particle filter.
Originally developed in meteorology and geosciences \cite{carrassi2018data}, data assimilation techniques have recently been employed in ABM research \cite{ward2016dynamic,lux2018estimation,clay2021real,cocucci2022inference}.
These works treat the ABM as a black box, adjusting ABM state variables so that forecasts come closer to observations.
The main advantage of data assimilation techniques over our approach is that they do not require building a new model (the learnable ABM).
At the same time, our approach offers several advantages.

\noindent
(i) It deals with the estimation of discrete variables in a natural and principled way. Standard data assimilation methods only allow to tune continuous variables \cite{ward2016dynamic,lux2018estimation}, and recent attempts to deal with discrete variables \cite{clay2021real,cocucci2022inference} tend to be heuristic and problem-specific.

\noindent
(ii) Its closed-form likelihood can be maximized with computationally efficient gradient-based methods, by leveraging deep learning frameworks/architectures.

\noindent
(iii) Such closed-form likelihood is also an essential tool to analyze identification problems, thus offering explanations about the estimated variables.

\noindent
(iv) While Kalman filters require Gaussian or quasi-Gaussian noise, and linear or weakly non-linear functional forms, our approach can easily integrate most types of stochastic element and non-linearities.

\noindent
Considering these advantages, we believe that likelihood-based estimation of ABM micro-states is a promising direction to obtain more principled approaches to data-driven ABMs.

\section*{Conclusions}
\label{sec:conclusions}

In this work, we have shown how to translate a complex agent-based model into a probabilistic graphical model to obtain a \emph{learnable} ABM.
For this type of model, methods such as maximum likelihood estimation can be used to estimate latent micro-state variables of the agents coherently with both the model and with provided data.
Then, we proposed an expectation-maximization algorithm for the resulting learnable ABM in order to estimate the latent variables given the observed ones.
We have shown that this process is indeed able to recover unobserved variables that are in line with the learnable model, as well as with the original one, under a variety of settings.
This way, we can feed such learned variables to the ABM, and obtain an evolution of its micro-states that is in line with the provided data.
This procedure empowers the ABM to be used as a forecasting tool.

Building a fine-grained link between an ABM and observed data opens the way for different exciting opportunities.
As we have shown in this work, it allows in the first place for better usage of ABMs as instruments for prediction.
Initializing agents' micro-states in a way that is coherent with observed data means that their future trajectory can be regarded as the best compromise between the theoretical model assumptions and the available observations.
Therefore, the quality of the predictions can also be used as a direct validation (or falsification) of the causal model embodied in the ABM.
Besides these immediate advantages, however, more advanced possibilities are opened.
For instance, defining the likelihood of the model w.r.t. the observations allows to perform model selection by using available data.
In other words, it allows using ABMs to formulate hypotheses and test them against real data.
While this technique has been shown to work properly in simple cases~\cite{monti2020learning}, its application to more complex ABMs requires further analysis and paves the way for novel research directions.
Furthermore, the translation of an ABM into a probabilistic model forces the modeler to lay bare their assumptions, and to consider the inferential problem.
This way, it brings forth possible identification problems: when different models (or realizations of the same model) lead to the same observable state, how can we choose one in practice?
Such a problem, often ignored in ABM research, will be vital to consider in further applications of ABMs to real-world data.

Our approach for learnable ABMs stems from the general framework of probabilistic graphical models~\citep{jordan2004graphical}.
While their application to ABMs opens possibilities for interdisciplinary cross-pollination, it also poses new theoretical challenges.
Because of the complexity of ABMs, many methods commonly applied such as Markov Chain Monte Carlo (MCMC)~\citep{jordan2004graphical} become computationally unfeasible.
Because ABMs aim to model emergent behavior through the combination of many simple rules, they often involve long chains of dependences among variables, often with highly non-linear behavior.
Hence, many desirable theoretical properties necessary for the convergence of MCMC might be missing, such as the uniqueness of the posterior distribution~\citep{hill2019stationarity}.
More in general, such a distribution is often very complex and high-dimensional, and difficult to learn through sampling techniques.
Thus, we choose to maximize the likelihood %
by leveraging gradient descent and automatic differentiation~\citep{van2018automatic}.
Interestingly, because of the long sequences of deterministic transformations typically found in ABMs, our optimization task ends up resembling deep learning ones. %
However, while the transformations in deep learning are purely data-driven---i.e., transformations aim only at maximizing prediction accuracy---our methodology still places an emphasis on causal mechanisms: each transformation represents an aspect of the theory being modeled.

\vfill{}
\clearpage
\section*{Materials \& Methods}
\footnotesize
\renewcommand{\thesubsection}{\Alph{subsection}}

\begin{table*}
\caption{Equations defining our agent-based model. \label{tab:model-equations}}
\fbox{
 \begin{minipage}{0.48\linewidth}
 \footnotesize
\begin{align*}
A_{t,x}&=A^I_{x}\frac{\sum_k M_{t-1,x,k}Y_k/N}{\sum_x \sum_k M_{t-1,x,k}Y_k/NL}\label{eq:M1}\tag{M1}\\
\pi_{t,x,k}&=\frac{\left(\min\left(0,Y_k-P_{t-1,x}\right)\right)^{1-\beta}A_{t,x}^\beta}{\sum_x \left[\left(\min\left(0,Y_k-P_{t-1,x}\right)\right)^{1-\beta}A_{t,x}^\beta\right]}\label{eq:M2}\tag{M2} \\
N^B_{t,x,k}&=Q\Gamma_k \pi_{t,x,k} \label{eq:M3}\tag{M3}\\
N^S_{t,x}&=R_{t-1,x}+\alpha(N-R_{t-1,x})\label{eq:M4}\tag{M4}\\
P^S_{t,x}&=P_{t-1,x}\left(1-\delta\left(1-\text{tanh }\left(\frac{\sum_k N^B_{t,x,k}}{N_{t,x}^S}\right)\right)\right) \label{eq:M5}\tag{M5}\\
D_{t,x}&=\min\left(\textstyle\sum_k N^B_{t,x,k},N^S_{t,x}\right) \label{eq:M6}\tag{M6}
\end{align*}
 \end{minipage}
 \begin{minipage}{0.48\linewidth}
 \footnotesize
\begin{align*}
\pi^D_{t,x,k}&=\frac{N^B_{t,x,k}\cdot \left(Y_k-P^S_{t,x}\right)}{\sum_{k'} \left( N^B_{t,x,k'}\cdot \left(Y_{k'}-P^S_{t,x}\right)\right) } \label{eq:M7}\tag{M7}\\
D^B_{t,x,k}&\sim\text{multinomial}\left(D_{t,x},\{\pi^D_{t,x,k}\}_k\right) \label{eq:M8}\tag{M8}\\
D^S_{t,x,k}&=D_{t,x}\frac{M_{t-1,x,k}}{\sum_{k'} M_{t-1,x,k'}} \label{eq:M9}\tag{M9}\\
P^B_{t,x}&= \sum_k \frac{Y_k D_{t,x,k}^B}{\sum_{k'} D_{t,x,k'}^B} \label{eq:M10}\tag{M10}\\
P_{t,x}&= \nu P^B_{t,x} + (1-\nu) P^S_{t,x} \label{eq:M11}\tag{M11} \\
M_{t,x,k}&= \max\left(0,M_{t-1,x,k}+D^B_{t,x,k}-D^S_{t,x,k}\right) \label{eq:M12}\tag{M12} \\
R_{t,x}&= R_{t-1,x} + N^S_{t,x} - D_{t,x} \label{eq:M13}\tag{M13}\\
\end{align*}
 \end{minipage}
}
\end{table*}

\subsection{Model description}
\label{sec:mmmodeldescr}

Here, we give a minimal description of the learnable model. In Supplementary \Cref{sec:detailed-learnable-abm} we provide a longer description as well as a detailed interpretation of each modeling assumption.

\begin{figure}[t!]
	\begin{tikzpicture}[scale=1.0]
		\scalebox{0.6}{
		\hspace{-14mm}
		\node[obs, minimum size=1cm] (Pprev) {$P_{t-1}$};
		\node[det, below=1.0cm of Pprev] (Mprev) {$M_{t-1}$};
		\node[det, right=6.7cm of Mprev, minimum size=1.25cm] (M) {$M_{t}$};
		
		\node[det, right=1cm of Pprev] (Nb) {$N^B_t$};
		
		\node[det, right=1cm of Mprev] (Ns) {$N^S_t$};
		\node[obs, right=1cm of Ns, minimum size=1cm] (Nd) {$D_t$};
		\node[det, right=1cm of Nd] (Ds) {$D^S_t$};
		
		\node[latent, right=3.1cm of Nb, minimum size=1cm] (Db) {$D^B_t$};
		\node[obs, right=0.8cm of Db, minimum size=1cm] (P) {$P_{t}$};
		\edge {Db} {P};

	  \node[obs, below=1.0cm of Mprev, minimum size=1cm] (Rprev) {$R_{t-1}$};
	  \node[det, below=1.0cm of M, minimum size=1.25cm] (R) {$R_{t}$};
		
		\node[det, right=3cm of Pprev] (Ps) {$P^S_t$};

		\node[latent, left=1.2cm of Mprev, minimum size=1cm] (M0) {$M_0$};

		\edge {Rprev} {Ns};
		\edge {Ns} {R};
		\edge {Nd} {R};
		
		\edge {Rprev} {R};
		
		\path[->]
			(Pprev) edge[bend left] node [left] {} (Ps);
		
		\edge {Pprev} {Nb};
		
		\edge {Mprev} {Ns};
		\edge[bend left] {Mprev} {Nb};
		
  \path[->]
    (Mprev) edge[bend right] node [left] {} (M);
		
		\path[->]
			(Nb) edge[bend left] node [left] {} (Db);
		\edge {Nb} {Nd};
		\edge {Nb} {Ps};
		
		\edge {Ns} {Nd};
		\edge {Ns} {Ps};
		
		\edge {Nd} {Ds};
		\edge {Nd} {Db};
		
		\edge {Ps} {Db};
		\path[->]
			(Ps) edge[bend left] node [left] {} (P);
		
		\edge {Db} {M};
		
		\edge {Ds} {M};
		
		\edge {M0} {Mprev};
		
		}
	\end{tikzpicture}
	\vspace{-20mm}
	\caption{Graphical model diagram of the learnable ABM for a time step $t$. See Materials \& Methods \Cref{sec:mmmodeldescr} for notation. Diamonds indicate deterministic variables, white circles indicate latent stochastic variables, grey circles indicate observed stochastic variables.}
	\label{fig:graphical-model}
\end{figure}
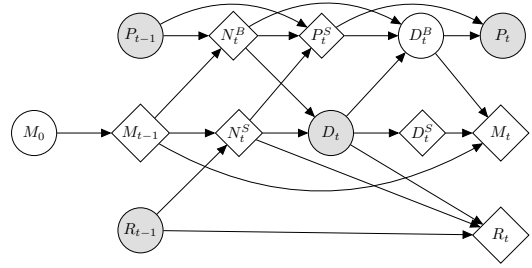

The model represents the housing market of a city with $L$ locations or neighborhoods denoted by $x=1,\ldots,L$, each with $N$ indistinguishable homes, inhabited by agents that are only distinguished by their income class $k=1,\ldots,K$.
The vector of state variables $Z_t$ is composed by the variables 
$\{M_{t,x,k}\},\{P_{t,x}\},\{R_{t,x}\}$, 
where $M_{t,x,k}$ is the number of agents of income $k$ living in location $x$ at time $t$;
$P_{t,x}$ is the average price of location $x$ at time $t$;
and $R_{t,x}$ is the inventory of unsold homes at location $x$ at time $t$. 
These state variables are updated according to deterministic and stochastic equations that represent the demand and supply sides of the housing market, and the matching between potential buyers and sellers.
The causal links between these variables are summarized in \Cref{fig:graphical-model}.
All the equations of the model, \Crefrange{eq:M1}{eq:M13}, are listed in \Cref{tab:model-equations}.

\Crefrange{eq:M1}{eq:M3} characterize the number of buyers from each income class that try to buy a house at each location at time $t$.
Buyers prefer to live in locations with higher attractiveness $A_{t,x}$, which depends on a constant local intrinsic attractiveness $A_x^I$ and on the time-varying average income at that location, captured through $Y_k$---the income of agents in income class $k$ \labelcref{eq:M1}.
However, locations with high attractiveness may also be more expensive, so the probability $\pi_{t,x,k}$ for a buyer of income class $k$ to choose location $x$ also depends on the difference between their possibility to pay---here exemplified by income $Y_k$---and price $P_{t,x}$ \labelcref{eq:M2}.
Finally, the number of potential buyers of income class $k$ for location $x$ at time $t$, $N^B_{t,x,k}$, is given by simply assuming that, for each income class $k$, a fraction $\Gamma_k$  of the total buyers $Q$ distribute themselves among all locations following probabilities $\pi_{t,x,k}$ \labelcref{eq:M3}.

Next, \Crefrange{eq:M4}{eq:M5} characterize the supply side of the market.
The number of sellers $N_{t,x}^S$ is given by the inventory of houses on sale at the previous time, $R_{t-1,x}$, plus a fraction $\alpha$ of the houses that were not on sale \labelcref{eq:M4}.
Moreover, the minimum price that the sellers at location $x$ are willing to accept, $P^S_{t,x}$, is a smooth function of the ratio between the number of buyers and sellers at $x$: when there are more buyers than sellers, sellers refuse to sell at a price below $P_{t-1,x}$; conversely, when there are more sellers, they are willing to sell at a discount, up to a price that is $1-\delta$ of $P_{t-1,x}$ \labelcref{eq:M5}.

The demand and supply sides of the market are matched in \Crefrange{eq:M6}{eq:M11}.
The number of deals or transactions $D_{t,x}$ is the short side of the market, i.e., the minimum between the number of buyers and sellers \labelcref{eq:M6}.
When there are more buyers than sellers, only some buyers are able to secure a deal.
The probability that agents from income class $k$ secure a deal at $x$ is represented by $\pi_{t,x,k}^D$, which is proportional to the number of buyers from class $k$ and to their income \labelcref{eq:M7}.
The number of buyers of each income class who secure a deal, $D_{t,x,k}^B$, is given by the $D_{t,x}$ realizations of a multinomial with parameter $\pi_{t,x,k}^D$ \labelcref{eq:M8}.
In this way, the outcome of this random variable has to be consistent with $D_{t}$: the total number of buyers in each location $x$ is fixed to $D_{t,x}$;
for each location, the buyers are distributed among income classes according to $\pi_{t,x}^D$.
The number of sellers from class $k$ who manage to sell at location $x$  ($D_{t,x,k}^S$) is instead simply proportional to the fraction of $k$-agents living in location $x$ \labelcref{eq:M9}.
With \Crefrange{eq:M6}{eq:M9} having determined the income classes of buyers and sellers, \Cref{eq:M10,eq:M11} specify the average (observed) price of transactions $P_{t,x}$.
The model assumes that this is a weighted mean \labelcref{eq:M11} of the maximum price that the average buyer is willing to pay, $P^B_{t,x}$ \labelcref{eq:M10}, and of the minimum price that sellers are willing to accept, $P^S_{t,x}$ \labelcref{eq:M5}.

As a last step, we update the remaining state variables in \Cref{eq:M12,eq:M13}, simply by tracking the number of buyers and sellers in each class and location.

\subsection{Algorithm derivation}
\label{sec:likelihood}

Here, we provide a more detailed description of our algorithm.
Following our online assumption, its goal is to estimate latent variables at time $t$ by looking at observables at the same time step, and treating all the previously estimated variables as fixed.
Specifically, $D^B_0, \dots, D^B_{t-1}$, i.e. the buyers who previously relocated, and the corresponding sellers $D^B_0, \dots, D^B_{t-1}$, are fixed.
Therefore, the algorithm observes $P_t$ and $D_t$, and then it provides a new estimate of $D^B_t$, and update the estimate of $M_0$.
In fact, since previous $D^B$ and $D^S$ are fixed, $M_t$ is a deterministic function of the latent variable $M_0$
\begin{equation*}
  M_{t+1} = M_t + D^B_{t} - D^S_{t} =
            M_0 + \sum_{\tau = 0}^{t} ( D^B_{\tau} - D^S_{\tau} )
\end{equation*}
allowing us to treat $M_0$, and not $M_t$, as our latent variable.

Since our observed variables are in principle also a deterministic result of the others, we model their observed value as a noisy proxy of the value determined by the agent-based model rules.
Specifically, for prices we assume that we observe $\tilde{P}_t$, given by $\tilde{P}_t=P_t + \epsilon_P$, where the error $\epsilon_P$ is normally distributed and $P_t$ is the deterministic estimate of prices as computed by the model (see \eqref{eq:M11}).
Similarly, the number of observed deals $\tilde{D}_t$ will follow $\tilde{D}_t = D_t + \epsilon_D$.

Now, computing the likelihood of the observed prices $\tilde{P}_t$ requires knowledge of the latent variable $D^B_t$, that is, the distribution of buyers among classes and locations, which is a discrete outcome of a stochastic process dependent on our main latent variable $M_0$.
Therefore we resort to the (Generalized) Expectation Maximization algorithm. %
In this way, we alternate between evaluating the expectation of $D^B_t$ and updating our estimate of $M_0$ under the current estimate of expectation.
The latter can be performed with online gradient descent, since---once we fixed the probability of each possible outcome of $D^B_t$---what remains of the likelihood is a continuous and differentiable function of $M_0$.

First, observe that $\tilde{P}_t$ and $\tilde{D}_t$ are independent given $M_0$.
In fact, $D_t$ is fixed given $M_0$; the distribution of $D^B$ is also fixed, since it is determined from $D_t$ and $\pi^D$; and the error $\epsilon_D$ is drawn independently from the extraction of $D^B$ from such distribution and from $\epsilon_P$.
Therefore we can factorize the complete-data likelihood w.r.t. observed data $\mathbb{D} = \{ \tilde{P}_t, \tilde{D}_t \}$ as:
\begin{equation}
  \label{eq:likelihood}
  \prob(\mathbb{D} | M_0) =
  \prob(\tilde{P}_t | M_0) \cdot
  \prob(\tilde{D}_t | M_0).
\end{equation}
Since computing $\prob(\tilde{P}_t | M_0) = \sum_{D_b} \prob(D_b | M_0) \prob(\tilde{P}_t | D_b, M_0)$ without the knowledge of the latent variable $D^B_t$ would be unfeasible, we resort to the Generalized Expectation Maximization algorithm, alternating these two steps until convergence:

\begin{enumerate}
  \item First, we evaluate the expectation of $D^B_t$ given the rest of the variables. Given the set $\Omega$ of all possible values of $D^B_t$, for each $D^B_t \in \Omega$ we evaluate $$
  q(D^B_t) := \prob(D^B_t | M^*_0)
  $$ where $M^*_0$ is the current estimate of $M_0$.
  This probability is computed from \eqref{eq:suppM8}.
  \item Then, we update the estimate of $M_0$ in order to increase the likelihood from \Cref{eq:likelihood}, by increasing the auxiliary function %
  \begin{equation}
    \label{eq:q-function-unfact}
    \mathbb{Q}(M^*_0) :=
      \sum_{D^B_t \in \Omega}
        q(D^B_t) \, \log \prob(\tilde{P}_t, \tilde{D}_t, | D^B_t, M^*_0 )
  \end{equation}
\end{enumerate}

Noting that $\tilde{D}_t$ does not depend on $D^B_t$ (see \Cref{fig:graphical-model}), the last probability can be decomposed as
\begin{equation}
  \label{eq:log-likelihood-given-Db}
   \log \prob(\tilde{P}_t, \tilde{D}_t | D^B_t,M^*_0 ) =
  \log \prob(\tilde{P}_t | M^*_0, D^B_t) +
  \log \prob(\tilde{D}_t | M^*_0).
\end{equation}
These two elements are given by the gaussian distribution of the errors $\epsilon_P, \epsilon_D \sim \mathcal{N}(0, \sigma)$ ($\sigma$ being a hyper-parameter), between $P_t$ and $\tilde{P}_t$, and $D_t$ and $\tilde{D}_t$ respectively.
Note that $P$ and $D$ are a deterministic function of the latent variable $M_0$ we are optimizing, through the chain of deterministic equations of the ABM (\Cref{tab:model-equations}). %
The only free variable is in fact $M_0$, since the previous variables from time steps $t' < t$ are assumed to be fixed, and the value of $D^B$ is known for the assumption of EM.
Therefore, since all these deterministic functions are continuous and differentiable in the general case, it is easy to update $M^*_0$ ascending the gradient $\nabla_{M^*_0}\mathbb{Q}(M^*_0)$.
The complexity of computing this gradient is left to differentiable programming frameworks.

Nevertheless, this approach presents a problem: the set $\Omega$ of possible values for $D^B_t$, the matrix of numbers of actual buyers for each class and each location, is potentially huge (precisely, $\tbinom{n+k-1}{n}$ with $n$ buyers and $k$ classes).
We solve this problem with two considerations: first, we will show that $\tilde{P}_{t, x} \indep \tilde{P}_{t, y} | D_{t}, \pi^D_{t}$; second, we are not interested in all the possible values for the number of actual buyers: the behavior of a single agent is irrelevant with respect to the data we observe.
Let us analyze these two key points.

The first consideration stems from the the independence of outcome in different neighborhoods: \mbox{$D^B_{t, x} \indep D^B_{t, y} | D_{t}, \pi^D_{t}$}.
This fact follows naturally from \eqref{eq:M8}, since all locations are independently drawn.
As a consequence, also the probability of observed prices $\tilde{P}_{t, x}$ and $\tilde{P}_{t, y}$ are independent from each other for any two locations $x \neq y$, since \eqref{eq:M10} and \eqref{eq:M11} do not have any inter-location effect, and the observation noise $\epsilon_P$ is also independent across locations.
Therefore \mbox{$\tilde{P}_{t, x} \indep \tilde{P}_{t, y} | D_{t}, \pi^D_{t}$} and we can write:
\begin{equation}
  \label{eq:decompose}
  \prob(\tilde{P_{t}}|M_0, D^B_{t}) = \prod_{x} \prob(\tilde{P}_{t, x} | M_0, D^B_{t, x})
\end{equation}
Thus, we can factorize \Cref{eq:q-function-unfact} in a more practical way.
Let us call $\Omega_x$ the set of all possible values for $D^B_{t, x}$, given a location $x$.
Then, our algorithm becomes an iteration of the following two steps until convergence.
\begin{enumerate}
  \item Evaluate $\forall x \in \{1, \dots, L \}$ and $\forall D^B_{t, x} \in \Omega_x$: \begin{equation}
    \label{eq:q-db-l}
    q(D^B_{t, x}) := \prob(D^B_{t, x} | M^*_0).
  \end{equation}
Note that any two values in $\Omega_x$ are mutually exclusive, so $\sum_{D^B_{t, x} \in \Omega_x} q(D^B_{t, x}) = 1$ holds for all $x$.
  \item Update $M^*_0$ by ascending the gradient $\nabla_{M^*_0}\mathbb{Q}(M^*_0)$ of %
  \begin{equation*}
      \mathbb{Q}(M^*_0): =
        \sum_{x}
        \sum_{D^B_{t, x} \in \Omega_x}
          q(D^B_{t, x}) \,
          \log \prob(\tilde{P}_{t, x}, \tilde{D}_t | M^*_0, D^B_{t,x} )
  \end{equation*}
  \begin{equation*}
      =
      \log \prob(\tilde{D}_t | M^*_0) +
      \sum_{x}
      \sum_{D^B_{t, x} \in \Omega_x}
        q(D^B_{t, x}) \,
        \log \prob(\tilde{P}_t | M^*_0, D^B_{t,x})
  \end{equation*} because of Equations \ref{eq:log-likelihood-given-Db} and \ref{eq:decompose}.
\end{enumerate}

\smallskip

To define the set $\Omega_x$ we take advantage of the second key point: two different values of $D^B_{t, x}$ might be indistinguishable in practice given our data, if they differ only by a few agents.
Thus, instead of considering all the possible partitions of the integer $D_{x,t}$ in $K$ positive integers, we only consider their quotient for a given constant $s$ (i.e., 
$ \left \lfloor D_{x,t} / s \right \rfloor $): this can be thought of as the possible outcomes obtained by moving \emph{groups} of $s$ agents at a time.
Any difference below $s$ is considered negligible.
In practice, we set $s$ as a consequence of the available memory.
Given a maximum size for $|\Omega|$, we set a threshold for each $|\Omega_x|$ proportional to its original size $\tbinom{n+k-1}{n}$.
Here, we keep in consideration the effective number of classes $k \leq K$ that can afford a location, since both $P_{t-1}$ and $Y$ are assumed to be known at time~$t$.
After setting this threshold, we find the minimum $s$ s.t. $|\Omega_x|$ respects such threshold.

\section*{Acknowledgments}
M.P. acknowledges funding from the James S. Mc Donnell Foundation.

\scriptsize
\bibliographystyle{unsrt}
\bibliography{references}

\begin{thebibliography}{10}

\bibitem{wilensky2015introduction}
Uri Wilensky and William Rand.
\newblock {\em An introduction to agent-based modeling: modeling natural,
  social, and engineered complex systems with NetLogo}.
\newblock Mit Press, 2015.

\bibitem{railsback2019agent}
Steven~F Railsback and Volker Grimm.
\newblock {\em Agent-based and individual-based modeling: a practical
  introduction}.
\newblock Princeton University Press, 2019.

\bibitem{axelrod1997dissemination}
Robert Axelrod.
\newblock The dissemination of culture: A model with local convergence and
  global polarization.
\newblock {\em Journal of conflict resolution}, 41(2):203--226, 1997.

\bibitem{lux2018estimation}
Thomas Lux.
\newblock Estimation of agent-based models using sequential monte carlo
  methods.
\newblock {\em Journal of Economic Dynamics and Control}, 91:391--408, 2018.

\bibitem{gatti2020rising}
Domenico Delli~Gatti and Jakob Grazzini.
\newblock Rising to the challenge: Bayesian estimation and forecasting
  techniques for macroeconomic agent based models.
\newblock {\em Journal of Economic Behavior \& Organization}, 178:875--902,
  2020.

\bibitem{windrum2007empirical}
Paul Windrum, Giorgio Fagiolo, and Alessio Moneta.
\newblock Empirical validation of agent-based models: {Alternatives} and
  prospects.
\newblock {\em Journal of Artificial Societies and Social Simulation}, 10(2):8,
  2007.

\bibitem{deffuant2008agent}
Guillaume Deffuant, Sylvie Huet, and Sarah Skerratt.
\newblock An agent based model of agri-environmental measure diffusion: {What}
  for?
\newblock In {\em Agent {Based} {Modelling} in {Natural} {Resource}
  {Management}}, pages 55--73. INSISOC, 2008.

\bibitem{lee2015complexities}
Ju-Sung Lee, Tatiana Filatova, Arika Ligmann-Zielinska, Behrooz
  Hassani-Mahmooei, Forrest Stonedahl, Iris Lorscheid, Alexey Voinov, Gary
  Polhill, Zhanli Sun, and Dawn~C. Parker.
\newblock The {Complexities} of {Agent}-{Based} {Modeling} {Output} {Analysis}.
\newblock {\em Journal of Artificial Societies and Social Simulation}, 18(4):4,
  2015.

\bibitem{ward2016dynamic}
Jonathan~A Ward, Andrew~J Evans, and Nicolas~S Malleson.
\newblock Dynamic calibration of agent-based models using data assimilation.
\newblock {\em Royal Society Open Science}, 3(4):150703, 2016.

\bibitem{clay2021real}
Robert Clay, Jonathan~A Ward, Patricia Ternes, Le-Minh Kieu, and Nick Malleson.
\newblock Real-time agent-based crowd simulation with the reversible jump
  unscented kalman filter.
\newblock {\em Simulation Modelling Practice and Theory}, 113:102386, 2021.

\bibitem{cocucci2022inference}
Tadeo~Javier Cocucci, Manuel Pulido, Juan~Pablo Aparicio, Juan Ru{\'\i}z,
  Mario~Ignacio Simoy, and Santiago Rosa.
\newblock Inference in epidemiological agent-based models using ensemble-based
  data assimilation.
\newblock {\em Plos one}, 17(3):e0264892, 2022.

\bibitem{dempster1977maximum}
Arthur~P Dempster, Nan~M Laird, and Donald~B Rubin.
\newblock Maximum likelihood from incomplete data via the em algorithm.
\newblock {\em Journal of the Royal Statistical Society: Series B
  (Methodological)}, 39(1):1--22, 1977.

\bibitem{pangallo2019residential}
Marco Pangallo, Jean-Pierre Nadal, and Annick Vignes.
\newblock Residential income segregation: A behavioral model of the housing
  market.
\newblock {\em Journal of Economic Behavior \& Organization}, 159:15--35, 2019.

\bibitem{loberto2022whatdo}
Michele Loberto, Andrea Luciani, and Marco Pangallo.
\newblock What do online listings tell us about the housing market?
\newblock Forthcoming, \textit{International Journal of Central Banking}, 2022.

\bibitem{monti2020learning}
Corrado Monti, Gianmarco De~Francisci~Morales, and Francesco Bonchi.
\newblock Learning {Opinion} {Dynamics} from {Social} {Traces}.
\newblock In {\em {ACM}}, {KDD}, pages 764--773, 2020.

\bibitem{dyer2021approximate}
Joel Dyer, Patrick Cannon, and Sebastian~M Schmon.
\newblock Approximate bayesian computation with path signatures.
\newblock {\em arXiv preprint arXiv:2106.12555}, 2021.

\bibitem{venkatramanan2018using}
Srinivasan Venkatramanan, Bryan Lewis, Jiangzhuo Chen, Dave Higdon, Anil
  Vullikanti, and Madhav Marathe.
\newblock Using data-driven agent-based models for forecasting emerging
  infectious diseases.
\newblock {\em Epidemics}, 22:43--49, 2018.

\bibitem{poledna2020economic}
Sebastian Poledna, Michael~Gregor Miess, and Cars~H Hommes.
\newblock Economic forecasting with an agent-based model.
\newblock {\em Available at SSRN 3484768}, 2020.

\bibitem{aleta2020modelling}
Alberto Aleta, David Martin-Corral, Ana Pastore~y Piontti, Marco Ajelli, Maria
  Litvinova, Matteo Chinazzi, Natalie~E Dean, M~Elizabeth Halloran, Ira~M
  Longini~Jr, Stefano Merler, et~al.
\newblock Modelling the impact of testing, contact tracing and household
  quarantine on second waves of covid-19.
\newblock {\em Nature Human Behaviour}, 4(9):964--971, 2020.

\bibitem{geanakoplos2012getting}
John Geanakoplos, Robert Axtell, J~Doyne Farmer, Peter Howitt, Benjamin Conlee,
  Jonathan Goldstein, Matthew Hendrey, Nathan~M Palmer, and Chun-Yi Yang.
\newblock Getting at systemic risk via an agent-based model of the housing
  market.
\newblock {\em American Economic Review}, 102(3):53--58, 2012.

\bibitem{welling2008deterministic}
Max Welling, Chaitanya Chemudugunta, and Nathan Sutter.
\newblock Deterministic latent variable models and their pitfalls.
\newblock In {\em Proceedings of the 2008 SIAM International Conference on Data
  Mining}, pages 196--207. SIAM, 2008.

\bibitem{carrassi2018data}
Alberto Carrassi, Marc Bocquet, Laurent Bertino, and Geir Evensen.
\newblock Data assimilation in the geosciences: An overview of methods, issues,
  and perspectives.
\newblock {\em Wiley Interdisciplinary Reviews: Climate Change}, 9(5):e535,
  2018.

\bibitem{jordan2004graphical}
Michael~I Jordan et~al.
\newblock Graphical models.
\newblock {\em Statistical science}, 19(1):140--155, 2004.

\bibitem{hill2019stationarity}
Stacy~D Hill and James~C Spall.
\newblock Stationarity and convergence of the metropolis-hastings algorithm:
  Insights into theoretical aspects.
\newblock {\em IEEE Control Systems Magazine}, 39(1):56--67, 2019.

\bibitem{van2018automatic}
Bart van Merrienboer, Olivier Breuleux, Arnaud Bergeron, and Pascal Lamblin.
\newblock Automatic differentiation in ml: Where we are and where we should be
  going.
\newblock In {\em NeurIPS}, 2018.

\bibitem{artinger2016heuristic}
Florian~M Artinger and Gerd Gigerenzer.
\newblock Heuristic pricing in an uncertain market: Ecological and
  constructivist rationality.
\newblock {\em Available at SSRN 2938702}, 2016.

\end{thebibliography}
\clearpage

\onecolumn
\normalsize

{\LARGE\bfseries Supplementary Information}

\renewcommand{\thesection}{S\arabic{section}}
\renewcommand{\thefigure}{S\arabic{figure}}
\renewcommand{\thetable}{S\arabic{table}}
\renewcommand{\thealgorithm}{S\arabic{algorithm}}
\renewcommand{\thesubsection}{S\arabic{section}.\arabic{subsection}}

\section{Original vs. learnable ABM}
\label{sec:old-model-comparison}

This section provides more details about the original ABM (\Cref{sec:original-abm-pseudocode}), the learnable ABM (\Cref{sec:detailed-learnable-abm}), and the comparison between the two (\Cref{sec:abm-comparison}).

\subsection{The original ABM}
\label{sec:original-abm-pseudocode}

Pangallo et al.~\cite{pangallo2019residential} introduce an ABM that describes the housing market of a city.
It is beyond the scope of this section to fully repeat the description of the model and to justify each assumption, so we give a brief overview and report the pseudocode of the model (see \Cref{alg:full-abm-run} and \Cref{alg:full-abm-cda}).

The city has $N \cdot L$ inhabitants (see \Cref{table:notation_original_abm} for a summary of the notation used), with $Q$ buyers coming to the city every time step to purchase a home, divided between $\gamma_k$ buyers belonging to income class $k$, $k=1,\ldots K$, such that $Q=\sum_k \gamma_k$.
Each inhabitant $i$ is characterized by four state variables that can change over time $t$: state $s_{t,i}$ (buyer, housed, seller), reservation price $P_{t,i}^R$, location $x_{t,i}$, and a categorical income $Y_{t,i}$, that belongs to one of the $K$ income classes.
The city is composed of $L$ neighborhoods or locations $x$ that are distinguished by their intrinsic attractiveness $A_x^I$, social attractiveness $A^S_{t,x}$ (which depends on which agent inhabit the neighborhood), and market price $P_{t,x}$.

The model is initialized following some protocol to locate agents with different income in the city (e.g., uniformly at random, or following some predefined spatial distribution).
After that, at each time step $t$ some buyers come to the city to purchase a home, some housed agents decide to leave and sell their home, and buyers and sellers at each location are matched via a continuous double auction.

\Cref{alg:full-abm-run} details the operations that occur at each time step.
First, the model updates some location-specific variables which reflect the change in social composition that occurred in the previous time step (\crefrange{alg:1:bookkeep}{alg:1:bookkeep-end}).
These include updating the average income and social attractiveness at each location and then computing the utility for agents of a given income class at a given location.
Next, buyers choose a location where they try to purchase a home (\crefrange{alg:1:buyers}{alg:1:buyers-end}).
After that, housed agents may put their home on sale with probability $\alpha$ and set a reservation price by applying a markup $\mu$ to the market price of the location where they live (\crefrange{alg:1:sellers}{alg:1:sellers-end}).
Agents that decided to sell their home in previous time steps but were unsuccessful reduce their price by a factor $\lambda$ every $\tau$ time steps (\crefrange{alg:1:reservation}{alg:1:reservation-end}).
Finally, buyers and sellers are matched at each location via a continuous double auction, and successful buyers replace successful sellers (\crefrange{alg:1:book}{alg:1:book-end}).

\Cref{alg:full-abm-cda} details the continuous double auction process.
At each location, there is a set of buyers $\mathcal{B}_{t,x}$ and a set of sellers $\mathcal{S}_{t,x}$ (\crefrange{alg:2:agentsets}{alg:2:agentsets-end}).
If there are in fact either no buyers, or no sellers, or all reservation prices of the buyers are lower than the reservation prices of the sellers, no transaction takes place, and the market price does not get updated (\crefrange{alg:2:case-no-transactions}{alg:2:case-no-transactions-end}).
If instead at least one transaction can occur (\crefrange{alg:2:book}{alg:2:book-end}), the following process takes place.
First, one creates auxiliary lists of buyers and sellers (also known as \textit{logs}), $\mathcal{O}_{t,x}^B$ and $\mathcal{O}_{t,x}^S$ respectively, and fills them as agents are drawn uniformly at random from the common pool of buyers and sellers.
Every time that a buyer with a higher reservation price than a seller enters $\mathcal{O}_{t,x}^B$ (or a seller with a lower reservation price than a buyer enters $\mathcal{O}_{t,x}^S$), the buyer with highest reservation price is matched with the seller with lowest reservation price.
The price of the transaction is the weighted mean of the respective reservation prices, with weight given by a parameter $\nu$ capturing bargaining power, and this individual transaction price is added to the list $\mathcal{P}_{t,x}$.
Finally, the seller leaves the city and the buyer settles where the buyer was.
The market price is computed as the mean of the transaction prices.

\begin{table}[b!]
	\footnotesize
	\caption{Notation used by Pangallo et al.~\cite{pangallo2019residential}. The first block is indexes, the second block global parameters, the third block agent attributes, the fourth block location attributes.} \label{table:notation_original_abm}
	\begin{tabular}{cll}
	    \toprule
			Type & Variable &  Meaning \\
		  \midrule
		  \multirow{4}{*}{\rotatebox{90}{Indexes}}
		  &$t$ & Time \\
		  &$i,j,h,\iota$ & Agent \\
		  &$k,\kappa$ & Income class \\
		  &$x$ & Location \\
		  \midrule
		  \multirow{14}{*}{\rotatebox{90}{Global parameters}}
		  &$N$ & Number of agents/housing units at any location\\
		  &$L$ & Number of locations\\
		  &$K$ & Number of income classes \\
		  &$A^\text{I, max}$ & Maximum intrinsic attractiveness \\
		  &$R$ & Radius intrinsic attractiveness \\
		  &$P_0$ & Initial price that is the same at all locations \\
		  &$\beta$ & Preference for housing goods (vs. non-housing goods) \\
		  &$\alpha$ & Probability to put house on sale \\
		  &$\mu$ & Markup \\
		  &$\lambda$ & Reduction in reservation price if sale unsuccessful  \\
		  &$\tau$ & Time steps of unsuccessful sale needed to reduce reservation price \\
		  &$\nu$ & Bargaining parameter \\
		  &$Y_k$ & Income of class $k$ \\
		  &$\gamma_k$ & Number of incoming agents of class $k$ at any time step \\
		  \midrule
		  \multirow{5}{*}{\rotatebox{90}{Agent}}
		  &$s_{t,i}$ & State of agent $i$. $s_{t,i}=0$: Buyer.  $s_{t,i}=1$: Housed. $s_{t,i}=2$: Seller \\
		  &$x_{t,i}$ & Location where agent $i$ searches if $s_{t,i}=0$, otherwise location where it lives if $s_{t,i}=1,2$  \\
		  &$P_{t,i}^R$ & Buyer reservation price of agent $i$ if $s_{t,i}=0$, seller reservation price if $s_{t,i}=2$ \\ 
		  &$Y_{t,i}$ & Income of agent $i$ \\
		  &$t^S_i$ & Time when agent $i$ becomes a seller \\
		  \midrule
		  \multirow{7}{*}{\rotatebox{90}{Location}}
		  &$d_x$ & Distance of location $x$ to the center \\
		  &$\overline{Y}_x$ & Mean income at location $x$ \\
		  &$P_{t,x}$ & Price of location $x$ at time $t$ \\
		  &$A_{t,x}$ & Attractiveness of location $x$ at time $t$ \\
		  &$A_{x}^I$ & Intrinsic attractiveness of location $x$ \\
		  &$A_{t,x}^S$ & Social attractiveness of location $x$ at time $t$ \\
		  &$U_{t,x,k}$ & Utility for $k$-buyers at location $x$ at time $t$ \\
		  \bottomrule
	\end{tabular}
\end{table}

\begin{algorithm}[h!]
    \footnotesize
    \caption{Original model run at time step $t$}
    \label{alg:full-abm-run}
    \begin{flushleft}
    \algorithmicrequire \; Global parameters $N, L, K, \beta, \alpha, \mu, \lambda, \tau, \nu, \{Y_k, \gamma_k\}_{k=1}^K$, $\{A_x^I\}_{x=1}^L$\\
    Agent variables $\{s_{t-1,i},P_{t-1,i}^R,x_{t-1,i},Y_{t-1,i}\}_{i=1}^{N^\text{TOT}}$,
    location variables $\{P_{t-1,x}\}$ \\
    \algorithmicensure Agent variables $\{s_{t,i},P_{t,i}^R,x_{t,i},Y_{t,i}\}_{i=1}^{N^\text{TOT}}$,
    location variables $\{P_{t,x}\}$ \; 
    \begin{algorithmic}[1]
    	\State $\overline{Y}_t=  \sum_{i=1}^{N^\text{TOT}} Y_{t,i}/N^\text{TOT} $ \Comment{Compute average income over the city} \label{alg:1:bookkeep}
	\For{$x = 1,\ldots,L$}  \Comment{Update attractiveness and utility}
    	\State $\overline{Y}_{t,x}= \sum_{i \text{ s.t. } x_{t,i}=x} Y_{t,i}/N $ \Comment{Compute average income at location $x$}
		\State $A_{t,x}^S = \overline{Y}_{t,x}/\overline{Y}_t$
		\State $A_{t,x}=A_x^I \cdot A_{t,x}^S$
		\For{$k=1,\ldots K$}
			\State{$U_{t,x,k}=\begin{cases} \left(Y_k-P_{t-1,x}\right)^{1-\beta} \left(A_{t,x}\right)^\beta & \text{ if } Y_k-P_{t-1,x}>0 \\
					0 & \text{ if } Y_k-P_{t-1,x}\leq 0 \end{cases}$} 
		\EndFor	  
    \EndFor \label{alg:1:bookkeep-end}
\For{$x = 1,\ldots,L$} \label{alg:1:buyers}
\For{$k=1,\ldots K$}
\State{$\pi_{t,x,k}=\frac{U_{t,x,k}}{\sum_x' U_{x',k,t}}$}\Comment{Probability that $k$-buyers choose location $x$}  
\EndFor	  
\EndFor	  \label{alg:1:buyers-partial-end}
    \For{$k=1,\ldots,K$}\Comment{Create buyers and let them choose a location}\label{alg:1:buyers-buy}
    	\For{$i=N^\text{TOT}+\sum_{\kappa=1}^{k-1}\gamma_\kappa+1,\ldots,N^\text{TOT}+\sum_{\kappa=1}^{k-1}\gamma_\kappa+\gamma_k$}\Comment{Do not duplicate id $i$}
    		\State $s_{t,i}=0$ \Comment{Buyer}
			\State $Y_{t,i}=Y_k$
			\State $P^R_{t,i}=Y_k$ \Comment{Buyer reservation price $=$ income}
			\State $x_{t,i} \sim \text{Categorical}\left(\pi_{t,1,k},\ldots,\pi_{t,L,k}\right)$ \label{alg:1:buyers-categorical}
    	\EndFor
    \EndFor \label{alg:1:buyers-end}
    \For{$i \text{ s.t. } s_{t-1,i}=1$} \Comment{Housed agents put their home on sale with probability $\alpha$} \label{alg:1:sellers}
    \If{$\text{Bernoulli} \left(\alpha\right)=1$}
    	\State $s_{t,i}=2$ \Comment{Seller}
    	\State $t^S_i=t$
    	\State $P^R_{t,i}=(1+\mu)P_{x_{t,i},t-1}$ \Comment{Reservation price is a markup $\mu$ over previous market price}\label{alg:1:markup}
    \EndIf
    \EndFor \label{alg:1:sellers-end}
    \For{$i \text{ s.t. } s_{t,i}=2$} \Comment{Sellers update reservation price} \label{alg:1:reservation}
    	\If{ $t-t^S_i \neq 0 \And t-t^S_i \text{ mod } \tau=0$ } \Comment{Every $\tau$ time steps after $t^S_i$}
    	\State{$P^R_{t,i}=P^R_{t-1,i} \cdot \lambda$} \Comment{Reduce by factor $\lambda$}
    	\EndIf
    \EndFor \label{alg:1:reservation-end}
    \For{$x = 1,\ldots,L$} \label{alg:1:book}
    	\State{$\{s_{t,i},P_{t,i}^R,Y_{t,i}\}_{i \text{ s.t. } x_{t,i}=x }$, $P_{t,x}$ =\begin{equation*}
    	\text{Continuous Double Auction (Algorithm \ref{alg:full-abm-cda})} (\nu, \{s_{t,i},P_{t,i}^R,Y_{t,i}\}_{i \text{ s.t. } x_{t,i}=x }, P_{t-1,x}) 
    	\end{equation*}}
    \EndFor \label{alg:1:book-end}
    	\end{algorithmic}
  \end{flushleft}
\end{algorithm}

\begin{algorithm}[h!]
    \footnotesize
    \caption{Continuous double auction in the original ABM.}
    \label{alg:full-abm-cda}
    \begin{flushleft}
    \algorithmicrequire \; Parameter $\nu$, Agent variables $\{s_{t,i},P_{t,i}^R,Y_{t,i}\}_{i \text{ s.t. } x_{t,i}=x }$, location variable $P_{t-1,x}$ \\
    \algorithmicensure \;  Agent variables $\{s_{t,i},P_{t,i}^R,Y_{t,i}\}_{i \text{ s.t. } x_{t,i}=x }$, location variable $P_{t,x}$
    \begin{algorithmic}[1]
    \State{$\mathcal{B}_{t,x}=\{i \text{ s.t. } s_{t,i}=0 \}$} \Comment{Buyers}\label{alg:2:agentsets}
    	\State{$\mathcal S_{t,x}=\{i \text{ s.t. } s_{t,i}=2\}$} \Comment{Sellers}\label{alg:2:agentsets-end}
    	\If{$\mathcal{B}_{t,x}=\emptyset \text{ or } \mathcal{S}_{t,x}=\emptyset \text{ or } \max_{i \in \mathcal{B}_{t,x}}P^R_{t,i} <  \min_{i \in \mathcal{S}_{t,x}}P^R_{t,i} $}\label{alg:2:case-no-transactions}
    		\State{$P_{t,x}=P_{t-1,x}$}\Comment{No transactions, so no update}\label{alg:2:case-no-transactions-end}
    	\Else\label{alg:2:book}
    		\State{$\mathcal{O}_{t,x}^B, \mathcal{O}_{t,x}^S = \emptyset$}\Comment{Sets of buyers and sellers in the order book}
    		\State{$\mathcal{P}_{t,x} = \emptyset$} \Comment{Set containing the prices of individual transactions}
    		\For{$i \sim \text{Uniform}\left(\mathcal{B}_{t,x}\cup \mathcal{S}_{t,x}\right)$}\Comment{Draw uniformly at random without replacement}
    			\If{$i \in \mathcal{B}_{t,x}$}
    				\State{$\mathcal{O}_{t,x}^B\leftarrow i$}
    			\Else
    				\State{$\mathcal{O}_{t,x}^S\leftarrow i$}
    			\EndIf
    			\If{$\mathcal{O}_{t,x}^B\neq\emptyset \text{ and } \mathcal{O}_{t,x}^S\neq\emptyset \text{ and } \max_{\iota \in \mathcal{O}_{t,x}^B}P^R_{t,\iota} >  \min_{\iota \in \mathcal{O}_{t,x}^S}P^R_{t,\iota} $}
    			\State{$j=\text{argmax}_{\iota \in \mathcal{O}_{t,x}^B} P^R_{t,\iota}$}
    			\State{$h=\text{argmin}_{\iota \in \mathcal{O}_{t,x}^S} P^R_{t,\iota}$}
    			\State{Remove $j,h$ from $\mathcal{O}_{t,x}^B,\mathcal{O}_{t,x}^S$ respectively}
    			\State{$\mathcal{P}_{t,x} \leftarrow \nu P^R_{t,j} + (1-\nu) P^R_{t,h}$}
    			\State{$s_{t,h}=1, Y_{t,h}=Y_{t,j}$}\Comment{Agent $j$ replaces agent $h$ and becomes housed}
    			\EndIf
    		\EndFor
    	\State{$P_{t,x}=\text{mean}(\mathcal{P}_{t,x})$}
    	\EndIf\label{alg:2:book-end}
    \end{algorithmic}
  \end{flushleft}
\end{algorithm}

\FloatBarrier

\subsection{Detailed description of the learnable ABM}
\label{sec:detailed-learnable-abm}

This section gives a more detailed description of the learnable ABM than \Cref{sec:mmmodeldescr} in Materials \& Methods, and details the interpretation for each equation of the model.
Table \ref{table:notation} can be used as a reference for notation throughout this section (although some notation overlaps with that of \Cref{table:notation_original_abm}, there are a few differences and so we prefer to present the two tables as separate).

\begin{table}[t]
	\centering
	\footnotesize
	\caption{Notation for the learnable ABM.}\label{table:notation}
	\begin{tabular}{llp{8cm}}
	    \toprule
			Symbol & Set & Meaning \\
		  \midrule
			$K$ & $\mathbb{N}$ & Number of income classes \\
			$L$ & $\mathbb{N}$ & Number of locations  \\
			\addlinespace
			$N$ & $\mathbb{N}$ & Number of houses per location  \\
			$\beta$ & $\mathbb{R}[0,1]$ & Preference for attractiveness relative to affordability \\
			$Q$ & $\mathbb{N}$ & Total number of buyers at any time step \\
			$\alpha$ & $\mathbb{R}[0,1]$ & Probability to put house on sale \\
			$\delta$ & $\mathbb{R}[0,1]$ & Maximum reduction in seller reservation price \\
			$\nu$ & $\mathbb{R}[0,1]$ & Bargaining power of sellers \\
			\addlinespace
			$A^I$ & $\mathbb{R}^{L}$ & Intrinsic attractiveness \\
			$Y$ & $\mathbb{R}^K$ & Income\\ 
			$\Gamma$ & $\mathbb{R}^K$ & Fraction of buyers by income class \\ 
			\addlinespace
			$M_t$ & $\mathbb{R}^{L \times K}$ & Number of housed agents \\
			$A_t$ & $\mathbb{R}^{L}$ & Total attractiveness \\
			$P_t$ & $\mathbb{R}^L$ & Transaction price \\ 
			$\pi_t$ & $\mathbb{R}^{L \times K}[0,1]$ & Probability to choose a location \\
			$N^B_t$ & $\mathbb{R}^{L \times K}$ & Number of potential buyers\\ 
			$R_t$ & $\mathbb{R}^{L}$ & Inventory of properties on sale \\ 
			$N^S_t$ & $\mathbb{R}^{L}$ & Number of potential sellers  \\ 
			$P^S_t$ & $\mathbb{R}^L$ & Reservation price for sellers  \\
			$D_t$ & $\mathbb{N}^{L}$ & Number of transactions (Deals) that actually take place \\ 
			$\pi^D_t$ & $\mathbb{R}^{L \times K}[0,1]$ & Probability that an agent is selected among the buyers to conclude a deal \\
			$D^B_t$ & $\mathbb{N}^{L \times K}$ & Number of potential buyers that complete a transaction \\
			$D^S_t$ & $\mathbb{R}^{L \times K}$ & Number of potential sellers that complete a transaction \\
			\bottomrule
			\multicolumn{3}{p{12cm}}{We first indicate the parameters $K$ and $L$ that determine the size of the variables, next we indicate model-wide parameters (i.e., scalar quantities that are fixed in time) and location- or income class-specific parameters (size $L$ or size $K$), and finally variables that can be location-specific (size $L$) or location-income-specific (size $L\times K$). We further show which quantities are constrained to the unit interval $[0,1]$.}
	\end{tabular}
\end{table}

\subsubsection{General set-up}

Agents are divided into $K$ income classes, each characterized by income $Y_{k}$, $k=1,\ldots,K$.
All agents within the same income class, also named $k$-agents, are assumed to be identical and indistinguishable.
The city is composed of $L$ locations denoted by $x$.

\subsubsection{Demand}
Let $M_{t,x,k}$ be the number of inhabitants of class $k$ living at location $x$ at time $t$.
As shown in Table \ref{table:notation}, this number is a real rather than an integer.
We make this choice for computational reasons but, as we typically deal with large values of $M$, it does not substantially affect our results.
We assume that each location $x$ is characterized by an attractiveness $A_{t,x}$ that can change over time, given by
\begin{equation}
A_{t,x}=A^I_{x}\frac{\left(\sum_k M_{t-1,x,k}Y_k\right)}{\sum_x \left(\sum_k M_{t-1,x,k}Y_k\right)}.\label{eq:suppM1}\tag{M1}
\end{equation}
In the equation above, the first term $A^I_{x}$ is an intrinsic attractiveness that is fixed over the simulation.
It captures relatively permanent city features, such as amenities, schools, and public transport.
The other term captures an attractiveness towards wealthier neighborhoods that can vary in time.
It is defined by the mean, one-period lagged, income at location $x$, $\overline{Y}_{t-1,x}=\sum_k M_{t-1,x,k}Y_k/N$, divided by the mean income over the city $\overline{Y}_{t-1}=\sum_x \sum_k M_{t-1,x,k}Y_k/NL$.
Thus, location $x$ whose mean income is higher than average, i.e., $\overline{Y}_{t-1,x}>\overline{Y}_{t-1}$, is, \emph{ceteris paribus}, more attractive than locations whose mean income is lower than average.

In their decision to move to location $x$, agents in income class $k$ also take into account the affordability of location $x$, i.e., the difference between their willingness to pay, here simply captured by their income $Y_k$,\footnote{As discussed more at length in \cite{pangallo2019residential}, willingness to pay is proportional to income, so it would be enough to rescale incomes by a constant factor. Here, since we are not working with real-world data, we take this constant factor equal to unity.} and the average price at $t-1$, $P_{t-1,x}$.
The utility of $k$-agents for location $x$ is then given by\footnote{This is an indirect utility function, derived from a standard utility function in urban economics and from the saturation of the budget constraint.
A standard assumption for the utility function in urban economic models is a Cobb-Douglas in which agents mix between housing services, here exemplified by the attractiveness $A_{t,x}$, and a composite of non-housing goods and services $z_{k,t}$, i.e., $U_{t,x,k}=z_{k,t}^{1-\beta} A_{t,x}^\beta$.
Agents face a budget constraint $P_z z_{k,t} + \mathbb{E}[P_{t,x}] = Y_k$, where $P_z$ is the price of the non-housing composite and $\mathbb{E}[P_{t,x}]$ is the price that they expect to pay to buy a house at location $x$ at time $t$.
Because agents do not know the price $P_{t,x}$, which will be formed as a consequence of their decisions, they use the previous price $P_{t-1,x}$ as their expectation for $P_{t,x}$, i.e., $\mathbb{E}[P_{t,x}] = P_{t-1,x}$.
Renormalizing prices so that $P_z=1$ and solving for $z_{k,t}$ yields the expression in \Cref{eq:induti}.}
\begin{equation}
    V_{t,x,k}=
\begin{cases}
    \left(Y_k-P_{t-1,x}\right)^{1-\beta}A_{t,x}^\beta,& Y_k > P_{t-1,x},\\
    0,              & Y_k \leq P_{t-1,x}.
\end{cases}
\label{eq:induti}
\end{equation}
where $\beta\in (0,1)$ gives the relative weight of attractiveness relative to affordability.
When $\beta$ is close to 1, agents care little about affordability, while when $\beta$ is close to 0 the opposite holds.
If location $x$ is unaffordable ($Y_k \leq P_{t-1,x}$) then $V_{t,x,k}=0$.
Buyers in income class $k$ are willing to bid up to their income $Y_k$, i.e., their reservation price $P^B_{t,x,k}$ is equal to $Y_k$.

Summing up, $k$-agents looking to buy a house in the city evaluate a utility $V_{t,x,k}$ for all locations $x$.
They then choose a location $x$ where they try to buy a house with probability $\pi_{t,x,k}$ proportional to $V_{t,x,k}$, i.e.
\begin{equation}
\pi_{t,x,k}=\frac{\left(\min\left(0,Y_k-P_{t-1,x}\right)\right)^{1-\beta}A_{t,x}^\beta}{\sum_x \left[\left(\min\left(0,Y_k-P_{t-1,x}\right)\right)^{1-\beta}A_{t,x}^\beta\right]}.\label{eq:suppM2}\tag{M2}
\end{equation}

We assume that a total of $Q$ agents come to the city at each time step looking to buy a house, and that a share $\Gamma_k$ of these agents is in income class $k$.
The number of buyers of income class $k$ at location $x$ at time $t$, $N_{t,x,k}^B$ is given by
\begin{equation}
N^B_{t,x,k}=Q\Gamma_k \pi_{t,x,k}, \label{eq:suppM3}\tag{M3}
\end{equation}
i.e., it is the expected value of a multinomial with $Q\Gamma_k$ trials and a probability vector given by the $L$ values of $\pi_{t,x,k}$, for all locations $x$ (\eqref{eq:suppM3}).

\subsubsection{Supply}
\label{sec:supp_supply}

In each location $x$, the other side of the market is composed by sellers.
We do not distinguish the income class of potential sellers, in the sense that we just keep track of the total number of agents willing to sell their house at location $x$ at time $t$, $N_{t,x}^S$.
The total number of sellers is obtained by summing the number of agents who wanted to sell at the previous time steps but did not succeed, $R_{t-1,x}$, and the number of agents who decide to put their house on sale at $t$.
In turn, this number is given by a fixed fraction $\alpha$ of the agents that had not decided to sell before $t$, which represent the difference between the total number of agents residing at location $x$, $N_x$, and $R_{t-1,x}$.
In formula,
\begin{equation}
N^S_{t,x}=R_{t-1,x}+\alpha(N_x-R_{t-1,x}).\label{eq:suppM4}\tag{M4}
\end{equation}

The way sellers determine their reservation price, i.e., the minimum price they are willing to accept, is more sophisticated.
Here, we assume that sellers are not willing to accept any price below the average price at the previous time step, $P_{t-1,x}$, as long as there are more buyers than sellers.
This choice captures the idea that, in this situation (known as a ``sellers market''), sellers have more bargaining power than buyers.
Conversely, when there are more sellers than buyers, sellers compete for the few buyers by being aggressive in reducing their reservation price.
So, they are willing to accept offers that can be below $P_{t-1,x}$.
We let $\phi_{t,x}=\left(\sum_k N^B_{t,x,k}\right)/N_{t,x}^S$ denote the ratio between the number of potential buyers and of potential sellers at location $x$ and time $t$.
We then assume that sellers are willing to accept prices lower than $P_{t-1,x}$ by a fraction $\delta$ if there are more sellers than buyers, i.e., $\phi_{t,x}\rightarrow 0$.
Conversely, when there are more buyers than sellers ($\phi_{t,x}\rightarrow \infty$), sellers are not willing to go below $P_{t-1,x}$.
We interpolate between these extreme values of $\phi$ by assuming a hyperbolic tangent functional form, i.e., we assume that the sellers' reservation price $P^S_{t,x}$ is given by 
\begin{equation}
P^S_{t,x}=P_{t-1,x}(1-\delta(1-\text{tanh }((\sum_k N^B_{t,x,k})/N_{t,x}^S))) \label{eq:suppM5}\tag{M5}
\end{equation}
Here, we assume that sellers decide on their reservation price independently of their income class.

\subsubsection{Matching}

At this point, the two sides of the market have been completely characterized, as we know the number of buyers in each income class $N^B_{t,x,k}$, their reservation price $P^B_{t,x,k}$, the number of sellers $N^S_{t,x}$, and their reservation price $P^S_{t,x}$.
It remains to be determined how buyers and sellers are matched,t and how this matching impacts future prices and the social composition of neighborhoods.

To start, let $D_{t,x}$ denote the number of \emph{deals} that occur at location $x$ and time $t$, i.e., the number of transactions that effectively occur between buyers and sellers.
This number is given by the ``short side of the market'', i.e., by the minimum between the number of potential buyers and potential sellers:
\begin{equation}
D_{t,x}=\min\left(\sum_k N^B_{t,x,k},N^S_{t,x}\right) \label{eq:suppM6}\tag{M6}
\end{equation}
In case there are fewer deals than potential buyers, i.e., $D_{t,x}<\sum_k N^B_{t,x,k}$, we need to decide which potential buyers are successful in actually buying a house and which are not.

To do so, we assume that demand is satisfied on a pro-rata basis, although correcting the pro-rata assumption by making richer buyers more likely to secure a deal.
This assumption captures a bargaining process in a more tractable way than explicitly simulating an auction.
Thus, the probability that $k$-agents are able to secure a deal at location $x$ and time $t$, denoted as $\pi^D_{t,x,k}$, is proportional to the number of potential buyers in that class, $N^B_{t,x,k}$, multiplied by the difference between the reservation price of $k$-buyers and that of sellers, $Y_k-P^S_{t,x}$:
\begin{equation}
\pi^D_{t,x,k}=\frac{N^B_{t,x,k}\cdot \left(Y_k-P^S_{t,x}\right)}{\sum_{k'} \left( N^B_{t,x,k'}\cdot \left(Y_{k'}-P^S_{t,x}\right)\right) } \label{eq:suppM7}\tag{M7}.
\end{equation}

Then, we compute the number of actual buyers of class $k$ at time $t$ in location $x$ by a multinomial with $D_{t,x}$ trials and a parameter vector of length $k$ given by $\pi^D_{t,x,k}$:
\begin{equation}
D^B_{t,x,k}=\text{multinomial}\left(D_{t,x},\{\pi^D_{t,x,k}\}_k\right) \label{eq:suppM8}\tag{M8}
\end{equation}
We further compute the number of actual sellers $D^S_{t,x,k}$ by assuming that all agents living in location $x$ are equally likely to sell, and so the share of $k$-agents among the sellers is proportional to the share of $k$-agents among the inhabitants, $M_{t-1,x,k}/N$:
\begin{equation}
D^S_{t,x,k}=D_{t,x}\frac{M_{t-1,x,k}}{\sum_k' M_{t-1,x,k'}} \label{eq:suppM9}\tag{M9}
\end{equation}
As above, this simplification ensures tractability, as keeping track of the number of sellers in each class over time would substantially increase the dimensionality of the space of state variables.

Next, after we determine which buyers and which sellers are successful in securing a deal, we need to determine the price of the transactions.
First, we compute the average buyer reservation price $P^B_{t,x}$ as
\begin{equation}
P^B_{t,x}= \sum_k (Y_k D_{t,x,k}^B/\sum_{k'} D_{t,x,k'}^B) \label{eq:suppM10}\tag{M10}.
\end{equation}
Then, we assume that the average transaction price is a weighted average between the reservation price of buyers and that of sellers:
\begin{equation}
P_{t,x}= \nu P^B_{t,x} + (1-\nu) P^S_{t,x} \label{eq:suppM11}\tag{M11}
\end{equation}
Here, $\nu$ denotes the bargaining power of sellers as, the larger $\nu$, the higher the transaction price will be.\footnote{It is always $P^B_{t,x} > P^S_{t,x}$ because (i) $P^B_{t,x}>P_{t-1,x}$ (as only buyers whose reservation price is larger than $P_{t-1,x}$ come to location $x$, see \Cref{eq:induti}) and (ii) $P^S_{t,x}<P_{t-1,x}$, see \Cref{eq:suppM5}.}

\subsubsection{Update of state variables}

It only remains to update the stocks of inhabitants $M_{t,x,k}$ of each class $k$ in each location $x$ and of unsuccessful sellers $R_{t,x}$ in $x$.
For each class and each location, the number of inhabitants is given by the number of inhabitants on the previous time step, plus the buyers who secured a deal, minus the sellers who were able to sell:
\begin{equation}
M_{t,x,k}= \max\left(0,M_{t-1,x,k}+D^B_{t,x,k}-D^S_{t,x,k}\right). \label{eq:suppM12}\tag{M12}
\end{equation}
The number of unsuccessful sellers is obtained by summing the unsuccessful sellers on the previous time step to the number of sellers at $t$, subtracting the number of deals:
\begin{equation}
R_{t,x}= R_{t-1,x} + N^S_{t,x} - D_{t,x} \label{eq:suppM13}\tag{M13}
\end{equation}

\subsection{Comparison between the original ABM and the learnable ABM}
\label{sec:abm-comparison}

We need to perform several modifications in order to make the original ABM learnable (see \Cref{tab:comparison} for a summary).

As a general principle, while in the original ABM the state of the system was described by the variables of individual agents $i$, in the learnable ABM we only consider counts of how many agents are within each income class.
For instance, in the original ABM we keep track of the state $s_{t,i}$, income $Y_{t,i}$ and location $x_{t,i}$ of each agent $i$, while in the learnable ABM the variable $M_{t,x,k}$ counts how many agents of income $k$ are either housed or sellers at location $x$, the variable $R_{t,x,k}$ counts the sellers that were not successful in selling, and so on.
This general modification does not lose much information.
Indeed, the key heterogeneity that distinguishes agents in the original ABM is which income class they belong to, which is kept in the learnable ABM.
We discuss below some examples where considering agent counts makes (or does not make) some difference.
The rationale for this general principle is that we assume that we only observe aggregate data at the level of locations, and so we must be parsimonious with unobserved variables.

 \begin{table}[t]
 	\centering
	\footnotesize
	\caption{Comparison between original and learnable ABM. The first column reports the equation number of the learnable ABM (\Cref{tab:model-equations}). The second column reports the lines of code that perform the same operation in the original ABM (\Cref{alg:full-abm-run}). The third column explains the changes. For each change, we indicate whether it fulfills principles P1 and P2.}
 	\label{tab:comparison}
 		\begin{tabular}{l l p{8cm}}
 			\toprule
 			Learnable & Original & Changes \\
 			\midrule
 			All & All & Do not track individual agents, only consider agent counts (P1) \\
 			\ref{eq:suppM1} & \crefrange{alg:1:bookkeep}{alg:1:bookkeep-end} & Identical \\
 			\ref{eq:suppM2} & \crefrange{alg:1:buyers}{alg:1:buyers-partial-end} & Identical \\
 			\ref{eq:suppM3} & \crefrange{alg:1:buyers-buy}{alg:1:buyers-end} & Instead of repeated draws from categorical distribution, use expected value of multinomial distribution (P1, P2)		 \\
 			\ref{eq:suppM4} & \crefrange{alg:1:sellers}{alg:1:sellers-end}  & Instead of repeated draws from Bernoulli distribution, use expected value of binomial distribution (P1, P2)		 \\
 			\ref{eq:suppM5} & \cref{alg:1:markup}, \crefrange{alg:1:reservation}{alg:1:reservation-end} & \vspace{-1em}
			\begin{squishlist}
 			\item Do not track the prices asked by individual agents, assume instead a location-specific seller reservation price (P1).
 			\item Make the reservation price depend on the ratio between buyers and sellers, rather than discretely reducing it as sale is unsuccessful (P2).
			\end{squishlist}	\\[-1em]
 			\ref{eq:suppM6}-\ref{eq:suppM13} & \crefrange{alg:1:book}{alg:1:book-end}  & \vspace{-1em}
			\begin{squishlist}
 			\item Remove random ordering of buyers and sellers that is not observed (P1).
 			\item Remove argmax operations (P2).
			\end{squishlist} \\[-1em]
 			\bottomrule						
 		\end{tabular}
 \end{table}

We now discuss the modifications one by one.
First note that the probability for buyers to search for a home in a given location does not change from the original to the learnable ABM.
For instance, although the specifications look different, \Cref{eq:suppM1} and \crefrange{alg:1:bookkeep}{alg:1:bookkeep-end} in \Cref{alg:full-abm-run} are identical.
At the same time, \Cref{eq:suppM2} is just a shorthand for \crefrange{alg:1:buyers}{alg:1:buyers-partial-end}.

When it comes to choosing a specific location, the learnable ABM essentially assumes the expected value of the stochastic process used in the original ABM.
In the original ABM, individual agents belonging to a given income class $k$ select a given location $x$ by drawing from a categorical distribution---a multinomial distribution with one trial (\cref{alg:1:buyers-categorical}).
Because all buyers belonging to the same income class are identical and have the same probability to choose a given location, it is completely equivalent to consider a multinomial distribution with $\gamma_k$ trials.
Indeed, this is what the learnable ABM does, except it considers the expected value of this distribution (\Cref{eq:M3}).
This choice is, once again, to limit the amount of stochasticity: as the model does not observe potential buyers, it would have to estimate the realization of this variable, and this may create computational problems.

There is a similar difference in the computation of the number of sellers.
The original ABM simulates the decisions of sellers individually (\crefrange{alg:1:sellers}{alg:1:sellers-end}).
As soon as a given housed agent becomes a seller with probability $\alpha$, its state changes and that specific agent, from that point on, acts as a seller.
In the learnable ABM we treat sellers as undistinguishable, and so it is sufficient to compute new sellers by drawing from a binomial distribution (in fact, by taking the expected value of that distribution, as in \Cref{eq:suppM4}, again to limit the amount of stochasticity).

There is a more substantial difference in the way sellers determine their reservation price.
In the original ABM they follow an \textit{aspiration level heuristic} \cite{artinger2016heuristic}, i.e., sellers start from a markup on the market price (\cref{alg:1:markup}), and then they decrease their reservation price if they are unable to sell (\crefrange{alg:1:reservation}{alg:1:reservation-end}).
The outcome of this heuristic is that prices tend to be higher in locations with higher demand, as sellers do not need to decrease their initial price much.
We implicitly capture this dynamics in the specification of \Cref{eq:suppM5}, as discussed in \Cref{sec:supp_supply}.
However, according to our specification, all sellers in the same location have the same reservation price.
This choice allows to treat sellers as undistinguishable in the learnable ABM, differently from the original ABM.

The final main difference lies in the way buyers and sellers are matched.
Lacking information on individual transactions, we cannot write a computationally tractable likelihood by keeping the explicit representation of a continuous double auction (\crefrange{alg:1:book}{alg:1:book-end}).
So we try to keep its main features while using a more tractable form.
We achieve this by giving more probability to be matched to buyers with higher income (\Cref{eq:suppM7}) and by computing the market price as the weighted mean of the average buyer and seller price, as in the original ABM.
Note that the model keeps some stochasticity in the way the matching process works in the learnable ABM.
In particular, we observe the number of transactions, and the realization of the matching affects the evolution of the social composition at each location, which is a key property of the model we wish to preserve.

\section{Supplementary results}
\label{sec:suppresults}

In this section we discuss some additional results that were not shown in the main text. 

\subsection{Hyperparameter selection} 
\label{sec:hyperparams}

The learning algorithm described in \Cref{sec:likelihood} has a number of hyperparameters that must be set.
It is beyond the scope of this section to explore the performance of the learning algorithm for each combination of hyperparameters, so we only explain how we assign some hyperparameter values, and explicitly show the effect of two hyperparameters that we consider particularly important.

\begin{itemize}
\item Initial guess for $M_0$.
The first expectation step of our expectation-maximization algorithm requires an initial guess for $M_0$.
We take a uniform $M_0$ (i.e. a situation in which there is the same number of inhabitants of each income class in each location) to which we add some random noise.
The realization of this noise is the only difference between runs of the learning algorithm on the same trace (all the other steps are deterministic).
We experimented with a few options for the variance of the noise, finding that it did not have substantial impact on the results, and so simply decided to draw the noise from a standard Gaussian pdf in logarithmic space.
\item Expectation-Maximization (EM) parameters.
The EM algorithm keeps iterating the expectation and maximization steps until some variables converge or up to a maximum number of steps.
We use a 5\%  threshold to decide on convergence, and allow for a maximum of 100 steps.
Moreover, in the maximization step the gradient descent algorithm requires a learning rate and a number of learning steps (within each EM cycle).
We choose a learning rate of 0.001 and a maximum of 4 learning steps.
Overall, in our explorations the results were not particularly sensitive to the EM parameters.
\item Number of epochs.
In our preliminary experiments we found that going beyond 3 epochs only marginally increased accuracy, and the marginal gain kept decreasing with the number of epochs.
In light of this preliminary evidence, we fixed the number of epochs to 5.
\item Parameter $\delta$.
The parameter $\delta$ in \Cref{eq:suppM5} has no counterpart in the original ABM, in which the reservation price was set by individual sellers following an aspiration level heuristic.
Thus, one needs to choose a value of $\delta$ that makes time series generated by the learnable ABM as similar as possible to time series generated by the original ABM.
We experimented with a few values, noting that $\delta=0.06$ seemed to yield the most similar time series.
\item Other model parameters.
All other parameters of the learnable ABM have a counterpart in the original ABM, and so we select the same values for both.
These are: $L=5$, $K=3$, $N=1000$, $Q=500$, $\alpha=0.1$, $\nu=0.1$, $\beta=0.5$, $Y=[10,50,90]$, $\Gamma=[0.5,0.4,0.1]$.
\end{itemize}

This leaves two hyperparameters whose effect we want to explore.
\begin{itemize}
\item Number of $D^B$ samples.
As detailed in \Cref{sec:likelihood} of Materials \& Methods, our EM algorithm only considers a subset of all possible values of the latent variable whose likelihood is estimated in the expectation step.
If the number of samples is higher, we expect that the performance of the algorithm improves, but this also leads to increased computational cost.
We expect that increasing the number of $D^B$ samples has a similar effect as increasing the number of epochs.
However, because the sampling from all possible values of $D^B$ is an original contribution of this paper, it is interesting to explicitly explore how it affects the performance of the algorithm.
\item Standard deviation of the noise on observables, $\sigma_P$ and $\sigma_D$ for $\epsilon_P$ and $\epsilon_D$ respectively.
As detailed in \Cref{sec:likelihood}, $P_t$ and $D_t$ are deterministic functions of their ancestors, and so in order to obtain a likelihood we model their observed values as a noisy proxy of the deterministic values.
How close the observed values must be to the deterministic values is governed by a variance $\sigma$: the higher $\sigma$, the more observed values can be far from model values.
First, note that the scale of $\sigma$ does not matter: if both $\sigma_P$ and $\sigma_D$ are multiplied by the same number, the log-likelihood only shifts by a constant.
What matters is the relative value of $\sigma_P$ with respect to $\sigma_D$.
When $\sigma_P$ is larger than $\sigma_D$, the learning algorithm should give more importance to $D_t$, while in the opposite case it should give more importance to $P_t$.
To study this effect, we set $\sigma_P=1$ without loss of generality, and vary $\sigma_D$.
\end{itemize}

\begin{figure}[!h]
  \centering
      \centering
      \includegraphics[width=0.4\linewidth]{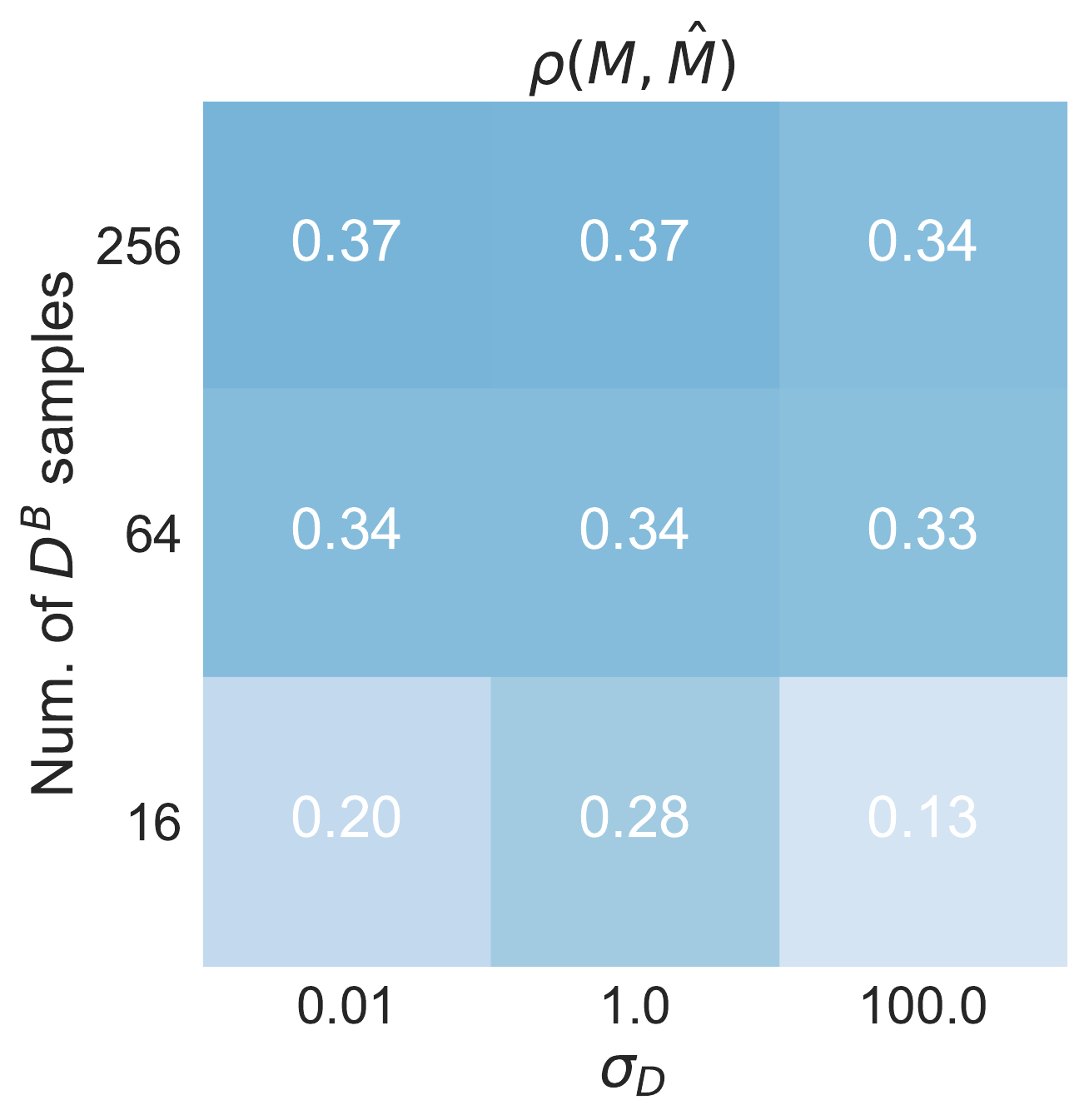} %
  \caption{\label{fig:hyperparam-pearson} %
  Hyperparameters selection for experiments with the original ABM, according to the Pearson correlation $\rho(M,\hat{M})$ between the ground truth $M$ and estimated values $\hat{M}$. %
  On the horizontal axis, we show the tested values for $\sigma_D$, which regulates the standard deviation of the prior distribution of the errors on the observed variable $D$. %
  On the vertical axis, we show the tested values for the number of considered samples of the stochastic variable $D^B$ (i.e. $|\Omega|$, described in Materials \& Methods). %
  }
\end{figure}

\Cref{fig:hyperparam-pearson} shows the Pearson correlation coefficient $\rho(M,\hat{M})$ between the ground-truth $M$ and the estimate $\hat{M}$ in the 10 traces that we use as a training set, for different choices of hyperparameters.
We consider three values of the number of $D^B$ samples, namely 16, 64, and 256, and three values of $\sigma_D$, namely 0.01, 1, and 100.

Increasing the number of $D^B$ samples monotonically increases $\rho(M,\hat{M})$, but the biggest gain is from 16 to 64 samples.
The situation when changing $\sigma_D$ is less clear: except for the case of 16 $D^B$ samples, there is little difference between $\sigma_D=0.01$ and $\sigma_D=1$, and a small degrade in performance when choosing $\sigma_D=100$.
These results suggest that $P_t$ and $D_t$ carry similar information for estimating the latent variables, and so giving more importance to one variable over the other does not substantially change results (note that in our simulations $P_t$ and $D_t$ are on the same scale).
For simplicity, we select $\sigma_D=1$ for our experiments.

\subsection{Estimate of $M_0$ over time and epochs}
\label{sec:M0_epochs}

Recall that, thanks to our online learning assumption, at each time step $t$ the algorithm estimates $M_0$ while keeping all changes in the number of inhabitants in previous steps fixed.
At the beginning of a new epoch, EM estimates $M_0$ starting from the latest estimate of $M_0$ in the previous epoch, and then repeats the same operations.
A wildly changing estimate of $M_0$, both within the same epoch and across epochs, would indicate that the model cannot converge on an estimate.

\begin{figure}[!h]
  \centering
  \includegraphics[width=0.7\linewidth]{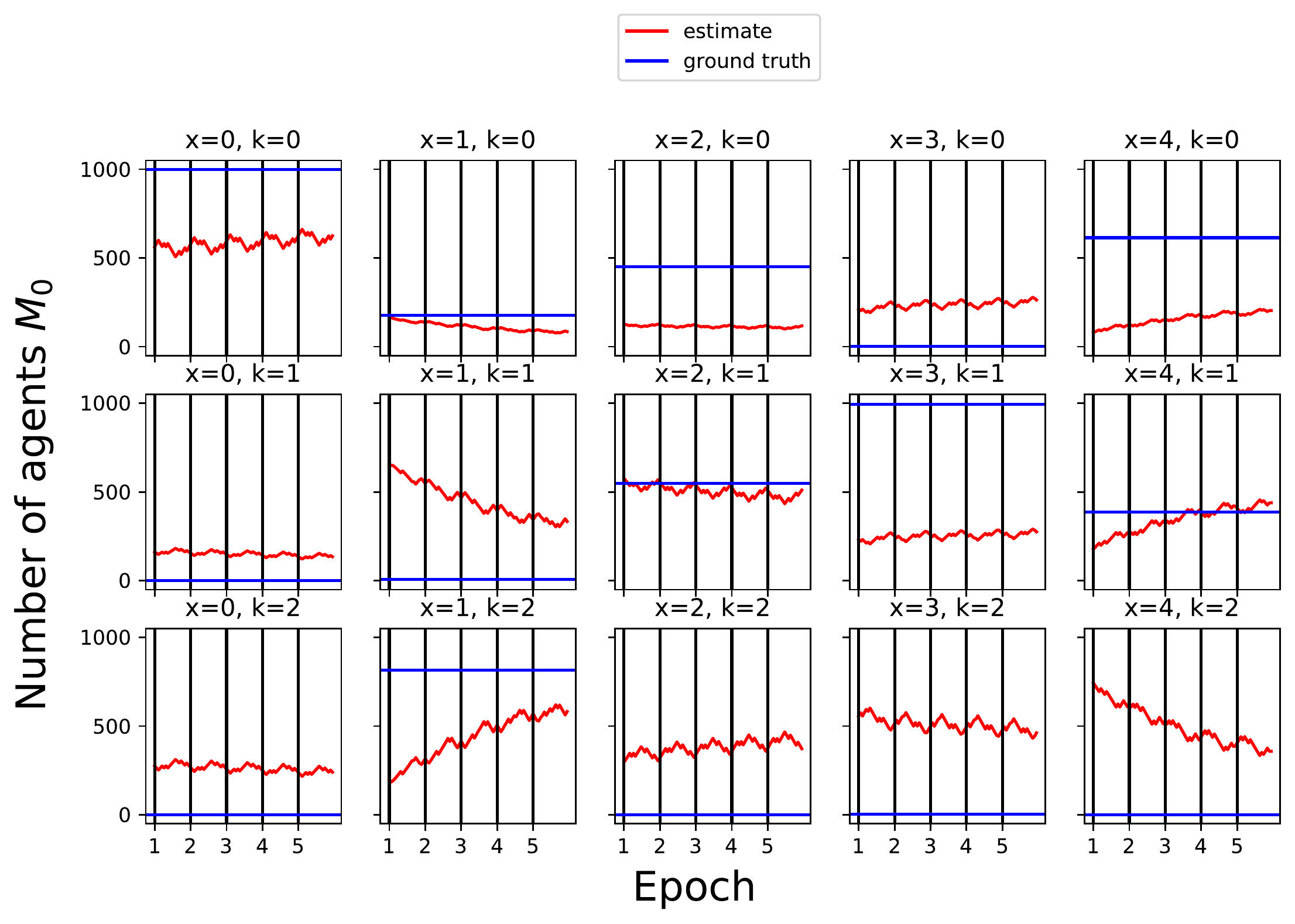}
  \caption{\label{fig:estimate_M0_multiple_epochs} %
  Estimate of $M_{0,x,k}$ over time and over multiple epochs, for the same simulation as the one shown in \Cref{fig:population-class}. Each panel represents a value of location $x$ and income class $k$.Each horizontal blue line is the ground truth value of $M_0$, each red line is the estimate of $M_0$ at a particular time step at a particular epoch. Black vertical lines distinguish the 5 epochs on which we run our experiments, within each epoch we estimate $M_0$ at each of 19 time steps.}
\end{figure}

\Cref{fig:estimate_M0_multiple_epochs} considers the same simulation as the one shown in \Cref{fig:population-class}.
It shows that the estimate of $M_0$ is actually relatively stable, for most locations and income classes.
For instance, in the case of locations $x=0,2,3$ there are no noticeable trends and the algorithm remains stuck in a local minimum that does not correspond exactly to the ground truth (although it still has a good correlation).
In locations $x=1,4$ there appears to be a trend that moves the estimation of $M_0$ closer to ground truth values (except for the case $x=1,k=0$), but convergence seems slow and, in our view, does not warrant increasing the number of epochs.

\subsection{Exploration of the loss}
\label{sec:loss}

Here we explore in detail how the loss $\mathcal L = -\log \prob(\mathbb{D} | M_0)$ depends on $M_0$ in three settings, shedding light into the performance of the algorithm.

\begin{figure}[!h]
        \centering
        \begin{subfigure}[b]{0.48\textwidth}
                \includegraphics[width=\textwidth]{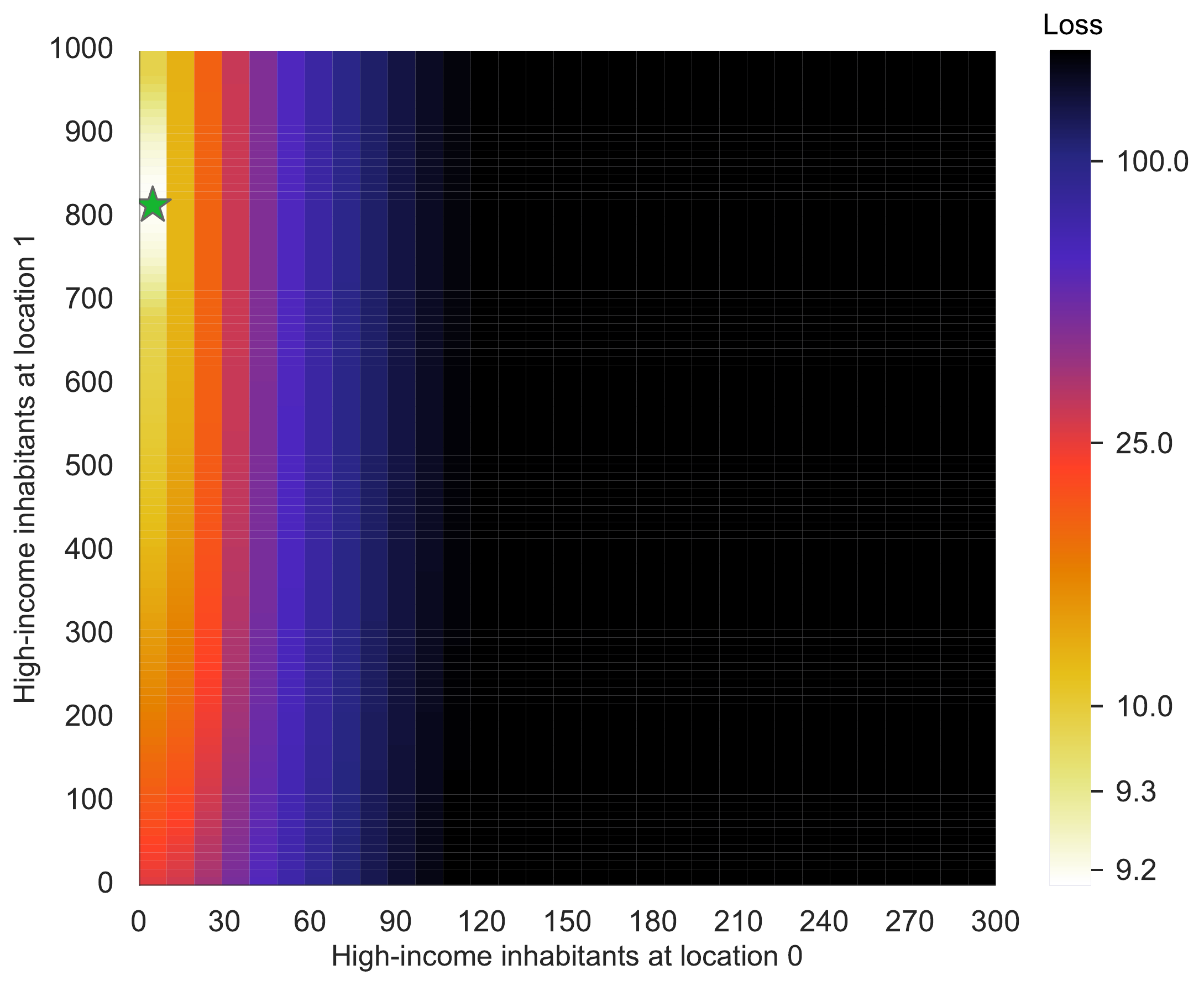}
								\caption{Learnable ABM as ground truth}
                \label{fig:heatmap-5-locs-learnable}
        \end{subfigure}
        ~ 
        \begin{subfigure}[b]{0.48\textwidth}
                \includegraphics[width=\textwidth]{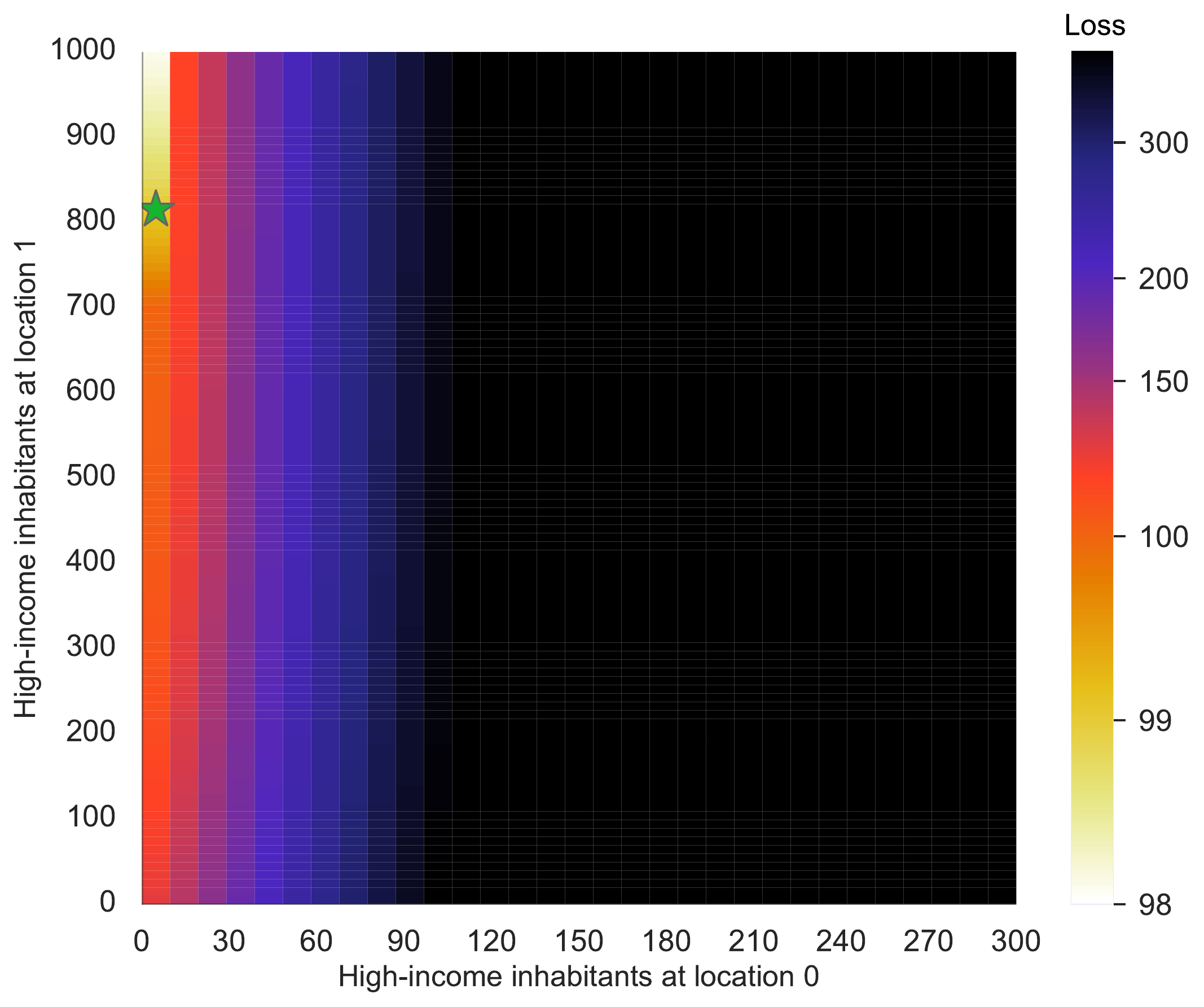}
								\caption{Original ABM as ground truth}
                \label{fig:heatmap-5-locs-original}
				\end{subfigure}        
				\caption{Heatmap of the loss (i.e., negative log likelihood) as a function of latent variables $M_{t=0, x=0, k=2}$ and $M_{t=0, x=1, k=2}$, representing the number of high-income inhabitants at two locations. The green star represents the ground truth value for these two variables. On the x axis we omit values beyond $300$ since their loss is much larger, for better presentation.}\label{fig:heatmap-5-locs}
\end{figure}

We first consider the same simulation as the one shown in \Cref{fig:population-class} and in the section above, and focus on time step 1, i.e. we take as observables $\mathbb{D} = \{ \tilde{P_1}, \tilde{D_1} \}$.
Further taking $D^B_1$ to be the same as the actual realization, we can explicitly compute the likelihood $\prob(\mathbb{D} | M_0)$ and hence the loss $\mathcal L$ as the opposite of the log-likelihood for several possible values of the variable we want to estimate, $M_0$.
In particular, we vary the number of inhabitants with highest income ($M_{0,x,k=2}$) at locations $x=0$ and $x=1$, ensuring that the total number of inhabitants at these locations remains equal to $N$.
To do so, we fix the number of middle income inhabitants ($M_{0,x,k=1}$) to their ground truth value, and fix the number of low-income inhabitants ($M_{0,x,k=0}$) to $M_{0,x,k=0}=N-M_{0,x,k=1}-M_{0,x,k=2}$.
These constraints make it possible to vary $M_{0,x=0,k=2}$ between 0 and 999 and $M_{0,x=1,k=2}$ between 0 and 992.
These values of $M_0$ include the ground truth, which corresponds to $M_{0,x=0,k=2}=0$ and $M_{0,x=1,k=2}=815$.
(All other components of $M_0$ are the same as in the ground truth.)

\Cref{fig:heatmap-5-locs} shows the loss as a function of these values of $M_0$, both taking the learnable ABM and the original ABM as ground truth
(for visualization purposes, we only show values of $M_{0,x=0,k=2}$ between 0 and 300, all other values lead to a much larger loss).
In the case of the learnable ABM (shown on the left) there is no mis-specification, and the loss attains its minimum possible value.
Since we take a Gaussian with unit variance and zero mean to model the error on $P_t$ and $D_t$, the minimum value of $\mathcal L$ is $\mathcal L = - \log \left( \sum_{x=0}^4 1/\sqrt{2\pi} \exp{(0)} \right) - \log \left( \sum_{x=0}^4 1/\sqrt{2\pi} \exp{(0)} \right)=-9.2$ when the errors $P_t-\tilde{P}_t$ and $D_t-\tilde{D}_t$ are zero.
As we see from the heatmap, the loss is well-behaved, in the sense that it does not display local minima and the minimum corresponds to the ground truth.

Interestingly, the gradient is much stronger when varying the number of high-income inhabitants at location 0, $M_{0,x=0,k=2}$, than when varying high-income inhabitants at location 1, $M_{0,x=1,k=2}$.
This effect can be explained by the initial price (not shown) at location 1 being much higher than the initial price at location 0, so that only the highest income agents can afford location 1 in the first place.
Therefore, making location 1 more or less attractive by increasing or decreasing the number of high-income agents inhabiting it does not make much of a difference to the distribution of buyers, compared to changing the number of high-income agents at location 0, which all the population can afford.

The right panel of \Cref{fig:heatmap-5-locs} takes the original ABM as the ground truth.
The heatmap is not much different, thus suggesting that there is no major misspecification.
However, the global minimum is a corner solution, at 1000 high-income inhabitants at location 1 and 0 high-income inhabitants at location 0, in contrast with the ground truth
(the ground truth is the same for the original and the learnable ABM).
Thus, while correctly guessing that location 1 has a much higher number of high-income inhabitants than location 0, the algorithm would not yield a perfect estimate.
This qualitative result is in line with the quantitative results that we show in \Cref{fig:estimates} of the main paper.

\begin{figure}[!h]
  \centering
  \includegraphics[width=1\linewidth]{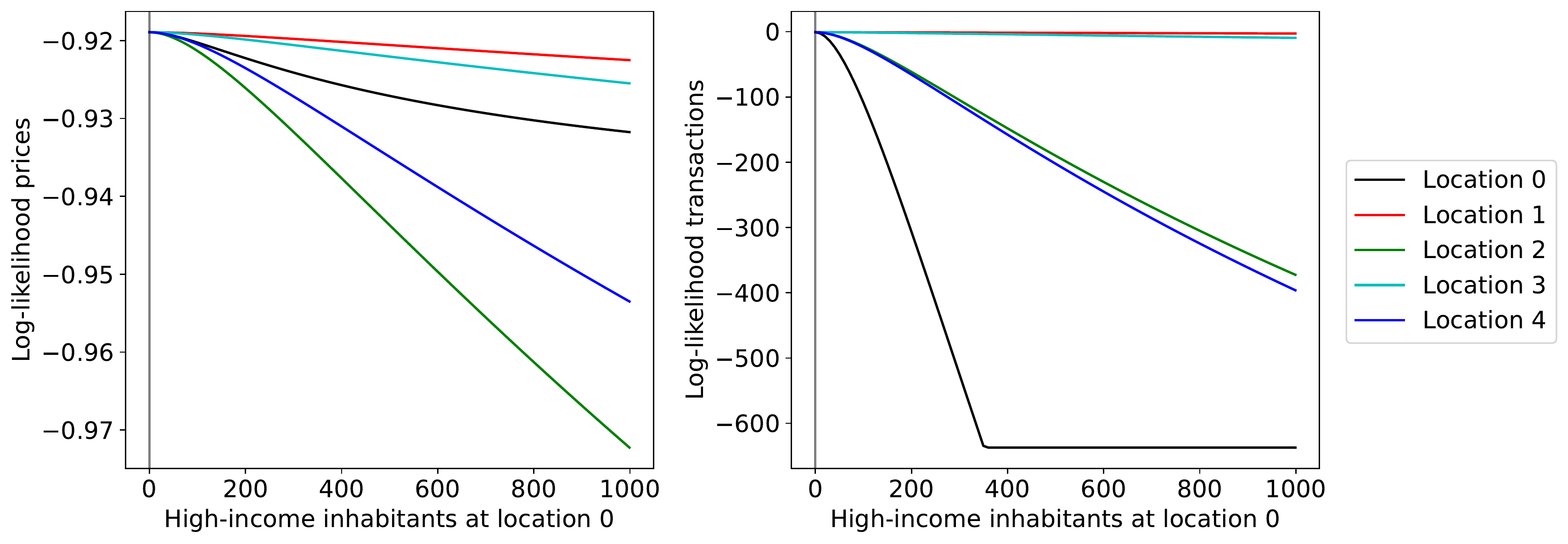}
  \caption{\label{fig:log_likelihood_5_locs} %
  Log-likelihood of prices and transactions at all five locations as we vary the number of high-income inhabitants at location 0. Here, the learnable ABM is the ground truth, and the ground-truth number of high-income inhabitants is represented as a vertical grey line.
  }
\end{figure}

Next, \Cref{fig:log_likelihood_5_locs} shows the log-likelihood of prices $\log \prob(\tilde{P_t} | M_0)$ and transactions $\log \prob(\tilde{D_t} | M_0)$ as a function of the number of high-income inhabitants at location 0, $M_{0,x=0,k=2}$, when taking the ground-truth value of the $M_{0,x=0,k=2}$ (and the learnable ABM is the ground truth).
Essentially, by taking the negative sum of all the components of the log-likelihood we obtain the loss, corresponding to a horizontal cut through the heatmap in \Cref{fig:heatmap-5-locs} (left) at the vertical coordinate of the ground truth.
The advantage of the representation in \Cref{fig:log_likelihood_5_locs} is that we can understand how each component contributes to the loss, and whether there are some non-linearities that give insights into difficulties to estimate the latent variables.

First, the log-likelihood varies much more with $D_t$ than with $P_t$.
As an intuitive justification for why this is the case, consider the graphical model in \Cref{fig:graphical-model}.
In that graphical representation, $P_t$ is influenced by $M_0$ only through several intermediate steps, while $D_t$ is more directly influenced.
In particular, changing the attractiveness of location 0 relative to the others changes demand across locations by a large margin, leading to very different values of $D_t$ at each location.
However, to propagate these differences to prices (at the same time step!) we go through $P^S_t$, which only varies by a factor $\delta=0.0625$ from the previous price, and so is not as sensitive to changes in $M_0$.

Second, the maximum of the log-likelihood is correctly achieved at $M_{0,x=0,k=2}=0$, and then all components of the log-likelihood monotonically decrease with $M_{0,x=0,k=2}$.
Focusing on the transactions (which dominate the loss), we see that the component of the likelihood corresponding to location $x=0$ is the most affected, the likelihood at locations 2 and 4 is also strongly affected, while the likelihood at 1 and 3 is barely affected.
The reason is that buyers at locations 1 and 3 are almost exclusively high-income, whereas locations 2 and 4 have many middle-income buyers (as location 0 does).
So, increasing the number of high-income buyers at location 0 strongly decreases the number of middle-income buyers at 2 and 4 (in a sense, these three locations get in competition), but it does not have a strong effect on locations 1 and 3.

Third, it is interesting that the log-likelihood for transactions at location 0 flattens out after a number of high-income inhabitatns $M_{0,x=0,k=2}=350$.
This effect is due to a supply constraint: increasing the number of high-income inhabitants at location 0 substantially increases demand, but the number of sellers is fixed, so, when the number of buyers becomes larger than the number of sellers, the number of transactions remains fixed (\Cref{eq:suppM6}).
Increasing the number of buyers still puts upward pressure on prices, and indeed the log-likelihood of prices does not flatten out.

\begin{figure}[!h]
  \centering
  \includegraphics[width=0.5\linewidth]{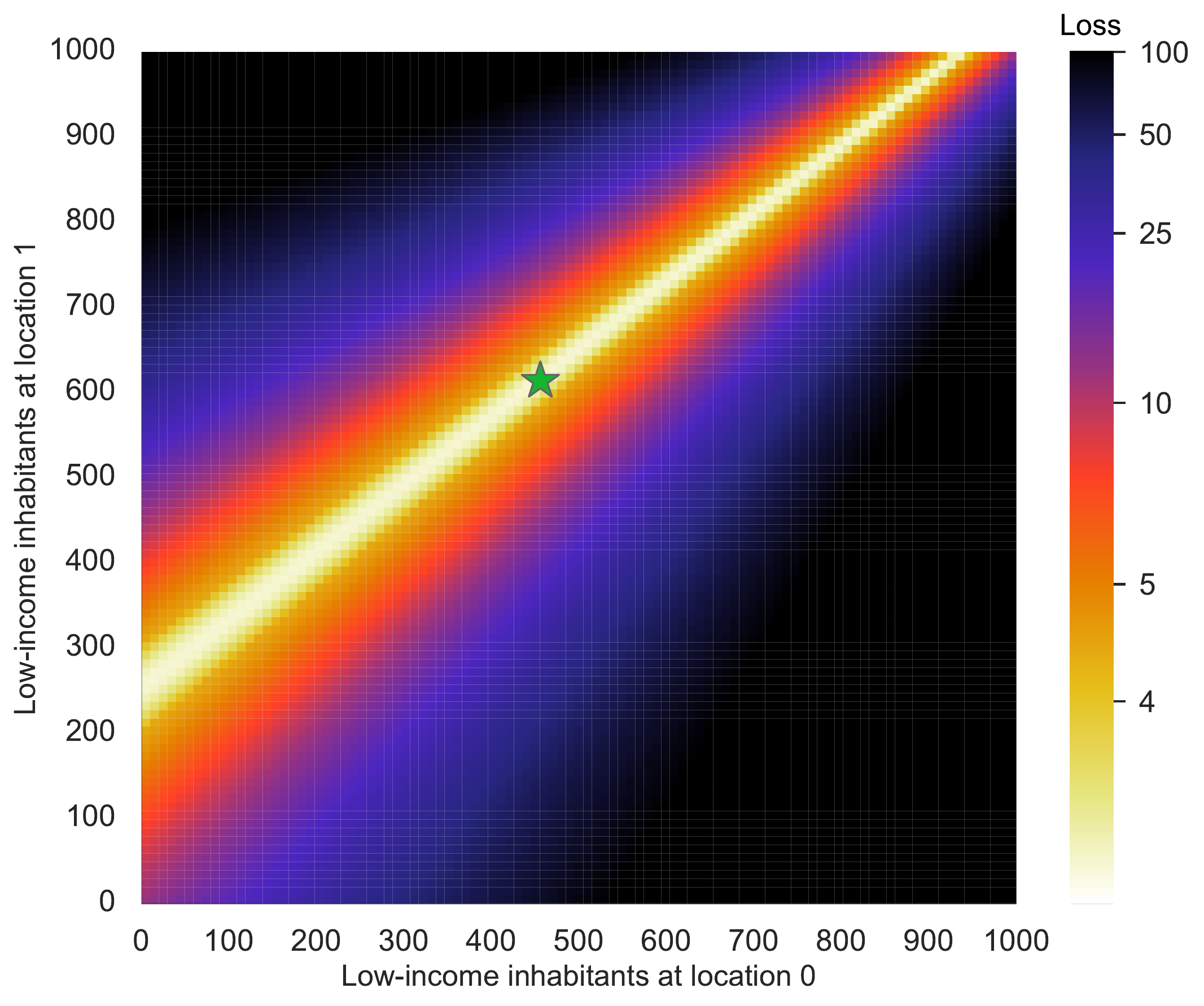}
  \caption{\label{fig:heatmap-2-locs} %
  Heatmap of the loss (i.e., negative log likelihood) as a function of latent variables $M_{t=0, x=0, k=0}$ and $M_{t=0, x=1, k=0}$, representing the number of low-income inhabitants at the two locations of this model. The green star represents the ground truth value for these two variables.
  }
\end{figure}

As a last example, we consider a simplified setting with $L=2$ locations, which allows to visualize a higher portion of the latent variable space than with $L=5$ locations.
Indeed, with $L=5$ locations we have 10 degrees of freedom in $M_0$ (considering $K=3$ and the constraint that the total number of inhabitants is $N$ at each location), but we can only visualize how the loss changes by varying 2 components and keeping all others fixed.
Instead, with $L=2$ locations we only have 4 degrees of freedom, so varying 2 components of $M_0$ at a time gives a more complete picture.

For simplicity, we initialize the model with the same parameters as above, focusing on the locations 2 and 4 above.
In this case, we have $M_{x=0,k,t=0}=[451,549,0]$ and $M_{x=1,k,t=0}=[614,386,0]$ as ground truth.
This time, we vary the number of low-income inhabitants $M_{0,x,k=1}$ at both locations, again using the same method as above to ensure that the total number of inhabitants is always $N$ at each location.

The results of this analysis are shown in \Cref{fig:heatmap-2-locs}.
Differently from \Cref{fig:heatmap-5-locs}, here the minimum of the loss is not attained at a single combination of values of $M_0$.
Instead, all points on a line that crosses the latent variable space from $M_{x=0,0,0},M_{x=1,0,0}=[0,250]$ to $[900,1000]$ lead to a value very close to the minimum loss (with only two locations, this value is  $\mathcal L = - \log \left( \sum_{x=0}^1 1/\sqrt{2\pi} \exp{(0)} \right) - \log \left( \sum_{x=0}^1 1/\sqrt{2\pi} \exp{(0)} \right)=-3.7$).
Intuitively, with just two locations, what matters is the relative attractiveness at one location compared to the other location.
So, as long as there are fewer low-income inhabitants at one location than at the other location, several possible configurations of $M_0$ lead to very similar values for the loss.

This situation constitutes an \textit{identification problem}: the model is not able to identify the ground truth, and any inference algorithm could converge on any value on the white line \Cref{fig:heatmap-2-locs}.
We conjecture that similar issues could prevent the learning algorithm from obtaining a perfect estimate for $M$.
Note that this problem is not due to the translation into a learnable form, but intrinsic to the ABM under scrutiny: many possible configurations of agents could lead to the same observable outcome.
Our approach allows to formally define and diagnose such issues, thus allowing ABM researchers to take into account the learnability of their model from observed data.

\end{document}